\documentclass[prd,showpacs,floatfix,twocolumn,amsmath,amssymb,floatfix,nofootinbib]{revtex4}
\usepackage{graphicx,color,dcolumn,booktabs,bm}
\usepackage{longtable,lscape}
\usepackage{txfonts}
\usepackage{overpic}
\usepackage{epsfig}
\usepackage{amssymb}
\usepackage{rotating}
\usepackage{epstopdf}
\usepackage{indentfirst}
\usepackage{feynmf}   
\usepackage{slashed}  
\usepackage{cases}
\usepackage{color}
\usepackage{multirow}
\usepackage{graphicx,color,dcolumn,booktabs,bm}

\graphicspath{{Figures/}} %

\graphicspath{{Figures/}} %
\usepackage[colorlinks, citecolor=blue,anchorcolor=red,menucolor=red, linkcolor=red,filecolor=red,runcolor=red,frenchlinks=red]{hyperref}
\usepackage{cases}
\begin{document}
\title{High-spin mesons below 3 GeV}
\author{Cheng-Qun Pang$^{1,2}$}{\email{pangchq13@lzu.edu.cn}
\author{Bo Wang$^{1,2}$}\email{wangb13@lzu.edu.cn}
\author{Xiang Liu$^{1,2}$\footnote{Corresponding author.}}\email{xiangliu@lzu.edu.cn}
\author{Takayuki Matsuki$^{3,4}$}\email{matsuki@tokyo - kasei.ac.jp}
\affiliation{$^1$School of Physical Science and Technology, Lanzhou University,
Lanzhou 730000, China\\
$^2$Research Center for Hadron and CSR Physics,
Lanzhou University $\&$ Institute of Modern Physics of CAS,
Lanzhou 730000, China\\
$^3$Tokyo Kasei University, 1-18-1 Kaga, Itabashi, Tokyo 173-8602, Japan\\
$^4$Theoretical Research Division, Nishina Center, RIKEN, Saitama 351-0198, Japan}
\begin{abstract}

In this work, we study the high-spin states with masses below 3 GeV observed in experiments and we perform analysis of mass spectrum and investigation of strong decay behaviors of the high-spin states. Comparing our results with the experimental data, we can reveal the underlying properties of these high-spin states; more importantly, we also predict their abundant decay features, which can provide valuable information for experimental exploration of these high-spin states.

\end{abstract}

\maketitle

\section{introduction}\label{sec1}

With the experimental progress of recent decades, the light-flavor meson family has become increasingly abundant (see the Particle Data Group (PDG) \cite{Agashe:2014kda}), there is a large amount of high-spin states with spin $J\geq 3$ available among the observed mesons listed in the PDG \cite{Agashe:2014kda}(see Table \ref{tab:1} for more details).
Although 26 high-spin states are collected in PDG, their properties are not presently well established. Therefore  it is necessary to determine how to categorize these high-spin states into meson families.

To provide a solution to this problem of studying high-spin states, we need to carry out a systematic and phenomenological investigation and combine it with the present experimental data.

For a $q\bar{q}$ meson system, the orbital quantum number of a meson is at least $L=2$, corresponding to the D-wave family when the spin quantum number is $J=3$. Thus, the high-spin states under discussion have close a relationship to $D$-wave, $F$-wave, {$G$-wave and  $H$-wave} meson families. 

\begin{table*}[htbp]
\caption{The observed high-spin states collected in the PDG \cite{Agashe:2014kda}. Here, the states listed as further states in the PDG are marked by the superscript $\mathfrak{f}$. The $C$ parity is valid only for the corresponding neutral states where the $J^{PC}$ quantum numbers of these high-spin states appear with isospin $I=1$. We need to emphasize that $ \rho_{3}(1690)$, $ \rho_{3}(1990) $, and $ \rho_{3}(2250)$ are not listed here since these three states have been studied in our former work \cite{He:2013ttg}. In the fifth column, we list some branching ratios and decay modes observed experimentally . {In this work we adopt the abbreviations, $\omega$, $\rho$, $\eta^\prime$, $a_0$, $b_1$,  $f_2$, and  $a_2$ for $\omega(782)$, $\rho(770)$, $\eta^\prime(958)$, $a_0(980)$, $b_1(1260)$,  $f_2(1270)$, and  $a_2(1320)$, respectively.}\label{tab:1}}
  \renewcommand{\arraystretch}{1.5}
 \tabcolsep=2pt
\begin{tabular}{lcccccccc}
\toprule[1pt]
 $I(J^{PC})$&{State} &                  {Mass (MeV)}&                           {Width (MeV)}   &      Other information        \\     
 \midrule[1pt]
 $0(3^{--})$&{$\omega_{3}(1670)$} &            {$1667\pm4$} &           {$168\pm 10$}  &       $ {\Gamma_{\omega\pi\pi}}/{\Gamma_{\rho\pi}}=0.71\pm0.27$ \cite{Diaz:1974bq}, $\pi b_1$ \cite{Baubillier:1979ve}\\                                                                                                                                           
$0(3^{--})$&{$\omega_{3}(1945)^\mathfrak{f}$} &                  {$1945\pm20$} &           {$115\pm22$}  & $\eta\omega$ \cite{Anisovich:2011sva}\\      
$0(3^{--})$&{$\omega_{3}(2255)^\mathfrak{f}$} &                  {$2255\pm15$} &           {$170\pm30$}  & $\eta\omega$  \cite{Bugg:2004rj, Anisovich:2011sva}\\      
$0(3^{--})$&{$\omega_{3}(2285)^\mathfrak{f}$} &                  {$2278\pm28$} &           {$224\pm50$}  &$\eta\omega$ \cite{Bugg:2004rj, Anisovich:2011sva}\\      
$0(3^{--})$&{$\phi_{3}(1850)$} &                  {$1854\pm7$} &           {$87^{+28}_{-23}$}  &$ {\Gamma_{KK^{*}}}/{\Gamma_{KK}}=0.55^{+0.85}_{  - 0.45}$ \cite{Aston:1988rf} \\
\hline
$0(3^{++})$&{$f_{3}(2050)^\mathfrak{f}$}&          {$2048\pm8$} &                       {$213\pm34$} &    $[\eta f_2]_{L=1, 3}, \pi a_2$  \cite{Anisovich:2000ut}\\

$0(3^{++})$&{$f_{3}(2300)^\mathfrak{f}$}&          {$2334\pm2$} &                    {$200\pm20$}  &     $[\eta f_2]_{L=1, 3}, \eta\prime f_2$       \cite{Bugg:2004rj}      \\     
$1(3^{++})$&{$a_{3}(1875)^\mathfrak{f}$}&          {$1874\pm43\pm96$} &            {$384\pm121\pm114 $}&                        $ {\Gamma_{f_{2}\pi}}/
                                                                                                                                        {\Gamma_{\rho\pi}}=0.8\pm0.2$,  ${\Gamma_{\rho_{3}
                                                                                                                                         (1690)\pi}}/{\Gamma_{\rho\pi}}=0.9\pm0.3$
                                                                                                                                         \cite{Chung:2002pu}\\     

 $1(3^{++})$&{$a_{3}(2030)^\mathfrak{f}$}&          {$2031\pm12$}&      {$150\pm18$}&     $\eta a_2$ , $\pi f_2$ \cite{Bugg:2004xu} \\     

$1(3^{++})$&{$a_{3}(2275)^\mathfrak{f}$} &            {$2275\pm35$} &           {$350^{+100}_{ - 50}$}  &$a_0 \eta$, $ \eta f_2$ \cite{Bugg:2004xu}
 \\
 \hline                                                                                                                                        $0(3^{+-})$&{$h_{3}(2025)^\mathfrak{f}$} &            {$2025\pm20$} &           {$145\pm 30$}  & $\eta\omega$  \cite{Anisovich:2011sva}\\        
$0(3^{+-})$&{$h_{3}(2275)^\mathfrak{f}$} &            {$2275\pm25$} &           {$190\pm 45$}  &  $\eta\omega$ \cite{Anisovich:2011sva}\\        
$1(3^{+-})$&{$b_{3}(2030)^\mathfrak{f}$} &            {$2032\pm12$} &           {$117\pm 11$}  &  $\omega\pi^0$ , $\pi^+\pi^-$  \cite{Anisovich:2002su}\\        
$1(3^{+-})$&{$b_{3}(2245)^\mathfrak{f}$} &            {$2245\pm50$} &           {$320\pm 70$}  &  $\omega a_2,~\omega\pi,~b_1\eta, ~\pi\omega(1650)$   \cite{Bugg:2004xu}\\        

\hline
$0(4^{++})$&{$f_{4}(2050)$}&             {$2018\pm 11$} &             {$237\pm18$}     & $ {\Gamma_{\omega\omega}}/                                                                                                                                        {\Gamma_{\pi\pi}}=1.5\pm0.3$ \cite{Alde:1990qd},                                                                                                                                     ${\Gamma_{\pi\pi}}/                                                                                                                                        {\Gamma_{\text{Total}}}=0.170\pm0.015$                                                                                                                                        \cite{Agashe:2014kda} \\                                                                                                                                        &&&&                                                                                                                                          ${\Gamma_{\eta\eta}}/ {\Gamma_{\text{Total}}}=(2.1\pm 0.8)\times 10^{ - 3}$  \cite{Alde:1985kp},                                                                                                                                            $                                                                                                                                       {\Gamma_{KK}}/{\Gamma_{\pi\pi}}=0.04^{+0.02}_{ - 0.01}$ \cite{Etkin:1981sg}\\
$0(4^{++})$&{$f_{4}(2300)$}&             {$2320\pm60$} &             {$250\pm80$}     &
$ {\Gamma_{\rho\rho}}/{\Gamma_{\omega\omega}}=2.8\pm0.5$ \cite{Barberis:2000kc}, $K^+K^-$ \cite{Cutts:1974vi},  $\pi\pi$ \cite{Hasan:1994he,Carter:1977jx}, $\eta\eta$  \cite{Amelin:2000nm},~$\eta f_2$   \cite{Bugg:2004xu}\\ 

$1(4^{++})$&{$a_{4}(2040)$}&            {$1996^{+10}_{-9}$} &                      {$255^{+28}_{ - 24}$}    &                                                                                                                                                $  {\Gamma_{\pi\rho}}/                                                                                                                                                {\Gamma_{\pi f_2}}=1.1\pm0.2\pm0.2$                                                                                                                                                \cite{Chung:2002pu}, $KK$ \cite{Cleland:1982te,Baldi:1978ju}, $\rho\omega$ \cite{Lu:2004yn}, $\eta\pi^0$ \cite{Uman:2006xb,Alde:1995fe,Anisovich:2001pn}, $\eta^\prime\pi$ \cite{Anisovich:2001pn, Ivanov:2001rv}   \\     
$1(4^{++})$&{$a_{4}(2255)^\mathfrak{f}$} &            {$2237\pm5$} &                   {$291\pm12$}  &  $\pi\eta$, $\pi\eta'$, $\pi f_2$  \cite{Bugg:2004xu}\\     
\hline$0(4^{-+})$&{$\eta_{4}(2330)^\mathfrak{f}$} &            {$2328\pm38$} &                   {$240\pm90$}  & {$a_0\pi$ \cite{Anisovich:2000ut}, $\pi a_2$ \cite{Anisovich:2000ut}, $\eta f_2$   \cite{Bugg:2004xu}  } \\     
 $1(4^{-+})$&{$\pi_{4}(2250)^\mathfrak{f}$} &            {$2250\pm15$} &                   {$215\pm25$}  &       { $a_0\eta$    \cite{Bugg:2004xu}}\\     
 \hline
  $0(4^{--})$&{$\omega_{4}(2250)^\mathfrak{f}$} &            {$2250\pm30$} &                   {$150\pm50$}  &       $\eta\omega$ \cite{Anisovich:2011sva}\\
$1(4^{--})$&{$\rho_{4}(2230)^\mathfrak{f}$} &            {$2230\pm25$} &                   {$210\pm30$}  &       $\omega\pi^0,~\pi^+\pi^-$  \cite{Anisovich:2002su}\\     
\hline

 $0(5^{--})$&{$\omega_{5}(2250) ^\mathfrak{f}$} &                      {$2250\pm70$} &                   {$320\pm95$}  &$\eta\omega$, $\pi b_1$  \cite{Bugg:2004xu}
                                                                                                                                       \\     

$1(5^{--})$&{$\rho_{5}(2350) $} &                     {$2330\pm35$} &                   {$260\pm70$}  & $\omega \pi^0$ \cite{Anisovich:2002su}, $\pi \pi$ \cite{Carter:1977jx, Hasan:1994he,Anisovich:2002su}, $K^+K^-$  \cite{Abrams:1967zzb, Alper:1980hi} \\
\hline

$0(6^{++})$&{$f_{6}(2510)$} &            {$2469\pm29$} &                   {$283\pm40$}  &         $ {\Gamma_{\pi\pi}}/{\Gamma_{\text{Total}}}=0.06\pm0.01$\cite{Binon:1983nx}

                                                                                                                                               \\     
 $1(6^{++})$&{$a_{6}(2450)$} &            {$2450\pm130$} &                   {$400\pm250$}  &      $KK$ \cite{Cleland:1982te}\\     
\bottomrule[1pt]
\end{tabular}
\end{table*}

In the following, we first focus on how to categorize these high-spin states into the conventional meson family, where the mass spectrum analysis is performed via the Regge trajectory. Furthermore, we calculate the Okubo-Zweig-Iizuka (OZI)-allowed two-body strong decay widths of these high-spin states, which can further test their possible meson assignments in combination with the present experimental data. We accordingly predict their abundant decay behaviors, which is important information for experimental exploration of high-spin mesons in future.

This paper is organized as follows. After a brief review,  we present in Sec. \ref{sec2} the experimental and theoretical research status. In Sec. \ref{sec3}, we adopt the Regge trajectory and the quark pair creation (QPC) model to study the high-spin states observed. 
The paper ends with conclusion and discussions in Sec. \ref{sec4}.

\section{Concise review of the present research status}\label{sec2}

Before illustrating our calculation, we first give a brief review of the research status of the high-spin states observed, which we hope is convenient for the readers.

\subsection{States with $J=3$}

An obvious peak signal was observed in the reaction $p\bar{p}\to \eta \pi^0\pi^0, \pi^0\pi^0, \pi^+\pi^-, \eta\eta$, and $\eta\eta\rq$ \cite{Anisovich:2000ut} which is named as $f_3(2050)$. Another $f_3$ state $f_3(2300)$ was introduced by performing the  partial wave analysis (PWA) of the data of $p\bar{p}\to\Lambda\bar{\Lambda}$ \cite{Bugg:2004rj}.
In Refs. { \cite{Anisovich:2000ut, Afonin:2007aa}}, $f_3(2050)$ was suggested to be a ground state of the $F$-wave meson family, with $f_3(2300)$ as its first radial excitation. In Ref. \cite{Masjuan:2012gc}, different from Ref. \cite{Anisovich:2000ut}, $f_3(2050)$ and $f_3(2300)$ were proposed as the first and the second radial excitations of the $f_3$ meson family, and an unobserved ground $f_3$ state with mass around $1.7$ GeV was predicted. 
{Reference. \cite{Afonin:2007aa}} obtained the same assignment as Ref. \cite{Anisovich:2000ut} by using the similar method to Ref. \cite{Masjuan:2012gc}. Ebert {\it et al}.  obtained the mass spectrum of some high-spin states via the relativistic quark model based on the quasipotential approach \cite{Ebert:2009ub}, where the mass spectrum calculation shows that $f_3(2300)$ can be a ground state in the $f_3$ meson family, which has dominant $s\bar{s}$ component.

There are three observed $a_3$ states. In the process $\pi^ -  p\to \pi^+\pi^ - \pi^ - p$, the E852 experiment reported a resonance $a_3(1875)$, and some ratios were measured. Two other states, $a_3(2030)$ and  $a_3(2275)$, were observed in the $p\bar{p}$ annihilation by SPEC \cite{Anisovich:2001pn,Anisovich:2001pp}.
In Ref. \cite{Masjuan:2012gc}, the authors suggested that $a_3(1875)$ and $a_3(2030)$ might be the same state, which could be a  ground state, with $a_3(2275)$ is the first radial excitation in the $a_3$ family. Additionally, the mass of the $n^{2s+1}L^J=1^3F_3$ state calculated by the relativistic quark model is 1910 MeV which corresponds to $a_3(1875)$ \cite {Ebert:2009ub}.

The SPEC experiment \cite{Anisovich:2011sva, Anisovich:2002su, Bugg:2004xu} observed $h_3(2025)$,  $h_3(2275)$, $b_3(2030)$, and  $b_3(2245)$  by analyzing the new Crystal Barrel data in the $p\bar{p}$ annihilation. These papers suggested that $h_3(2275)$ and $b_3(2245)$ are the first radial excitations of $h_3(2025)$ and $b_3(2030)$, which are the ground states in the $h_3$ and $b_3$ meson families, respectively. In Refs. \cite{Masjuan:2012gc,Afonin:2007aa}, it is suggested that that we regard $h_3(2025)/b_3(2030)$ and $h_3(2275)/b_3(2245)$ as the first and second
radial excitations in the $h_3/b_3$ meson families, where the corresponding ground states were predicted.
{
The $b_1(1640)$ state was predicted in Refs. \cite{Anisovich:2002us, Anisovich:2003tm}}. The study in Ref. \cite{Ebert:2009ub} indicates that $h_3(2275)$ and  $b_3(2245)$ could be the ground states with a component $s\bar{s}$.

The $\omega_3(1670)$ state was first found in the $\pi^+n \to p3\pi^0 $ process \cite{Armenise:1900zz} and has been studied by other experiments (the details for the experimental information on
$\omega_3(1670)$ are listed in the PDG \cite{Agashe:2014kda} ). $\omega_3(1670)$ can decay into $\rho \pi$ and $\pi\omega\pi$. The first radial excitation of $\omega_3$ is $\omega_3(1945)$, which was reported by SPEC \cite{Anisovich:2011sva}. At the same time, Ref.\cite{Anisovich:2011sva} also found $\omega_3(2255)$ and $\omega_3(2285)$, which were later confirmed by RVUE \cite{Bugg:2004rj} and $\omega_3(1670)$ was suggested to be the ground state of the $\omega_3$ family. Combining the PWA with the $n$-$M^2$ plot, the authors of Ref. \cite{Anisovich:2011sva} proposed that $\omega_3(1945)$ and $\omega_3(2285)$ are $2^3D_3$ and  $3^3D_3$ states, respectively \cite{Anisovich:2011sva}, while $\omega_3(2255)$ is a $^3G_3$ state. Reference  \cite{Afonin:2007aa} suggested $\omega_3(2285)$ could be the $1^3G_3$ state. The mass spectrum calculation  given in \cite {Ebert:2009ub} shows that   $\omega_3(1945)$  and $\omega_3(2285)$ could be the $1^3G_3$  and $2^3G_3$ states, respectively.

HBC  found $\phi_3(1850)$ in the $KK$ and $KK^*$ channels from the $K^-p$ collision  \cite{AlHarran:1981tk}, and it was confirmed by  OMEGA \cite{Armstrong:1982jj} and LASS \cite{Aston:1988rf}.
Both the $J$-$ M^2$ plot analysis in Refs. \cite{Anisovich:2002us, Anisovich:2003tm, Masjuan:2012gc} and the calculation of the mass spectrum  in \cite {Ebert:2009ub} show that $\phi_3(1850)$ is a good candidate for the $1^3D_3$ state.

\subsection{States with $J=4$}\label{sub-b}

The Serpukhov-CERN Collaboration \cite{Apel:1975at} and CERN-Munich Collaboration \cite{Blum:1975rk} observed a peak structure in the processes $\pi^-p \to n2\pi^0$ and $\pi^-p \to nK^+K^-$, which was named $f_4(2050)$. Other experiments
relevant to the observation of $f_4(2050)$ can be
found in PDG \cite{Agashe:2014kda}. Some observed channels and branching ratios are listed in Table \ref{tab:1}. CNTR \cite{Abrams:1969jm} first reported the resonance $f_4(2300)$; in the past decades, it has appeared in the reactions, $p\bar{p} \to K^+K^-$ \cite{Carter:1978ux}, $\pi\pi$ \cite{Carter:1977jx,Dulude:1978kt, Hasan:1994he}, $\eta\pi^0\pi^0$ \cite{Anisovich:2000ut},  and $\pi^-p \to K^+K^-$ \cite{Amelin:2000nm}.

There are some theoretical studies on the properties of the observed $f_4$ states.
The $f_4(2050)$ is treated as a molecule state composed of three $\rho$ mesons \cite{Roca:2010tf}.
Ebert {\it et al}. obtained a $1^3F_4$ $q\bar{q}$ state with mass $M = 2018 $ MeV and a $2^3F_4$ $q\bar{q}$ state with mass $M = 2284 $ MeV which correspond to $f_4(2050)$ and  $f_4(2300)$, respectively \cite {Ebert:2009ub}. Many studies  support this assignment \cite{Afonin:2007aa,Anisovich:2000kxa,  Anisovich:2002us, Anisovich:2003tm}.
 Additionally, in Ref. \cite{Masjuan:2012gc}, the Regge trajectory analysis shows that $f_4(2050)$
 and $f_4(2300)$ are the ground states dominated by
the $q\bar{q}$ and $s\bar{s}$ components, respectively.

There are many experiments relevant to $a_4(2040)$.
OMEGA observed a resonance with mass around 2030 MeV by the PWA of $\pi^-p \to n3\pi$ \cite{Corden:1977em}. Later,
$a_4(2040)$ was also found in the reactions,  $\pi p \to K_sK^{\pm}p$ \cite{Cleland:1982te}, $\pi^- p \to \eta \pi^0 n$ \cite{Alde:1995fe}, $\pi^- A \to \omega \pi^-\pi^0 A^*$ \cite{Amelin:1999gk}, $\pi^- p \to \eta' \pi^-p$ \cite{Ivanov:2001rv},
$\pi^- p \to \omega\pi^-\pi^0p$ \cite{Lu:2004yn}, and
$\pi^- P_b \to \omega\pi^-\pi^-\pi^+P_b'$ \cite{Lu:2004yn}. The observed decay modes are listed in Table \ref{tab:1}.SPEC also reported $a_4(2255)$ in $p\bar{p} \to  \pi^0\eta, 3\pi^0, \pi^0\eta\prime$ \cite{Anisovich:2001pn} and E835 confirmed the state
$a_4(2255)$ in the reaction $p\bar{p} \to \eta \eta\pi^0$ \cite{Uman:2006xb}.
All Regge trajectory studies show that $a_4(2040)$ is the ground state of the $a_4$ family, and $a_4(2255)$ is its first radial excitation \cite{Afonin:2007aa, Anisovich:2000kxa,Anisovich:2002us,Anisovich:2003tm, Masjuan:2012gc}. Ebert {\it et al}. obtained a $1^3H_4$ state with mass $M = 2234 $ MeV which is very close to $a_4(2255)$ \cite {Ebert:2009ub}. However, in Ref. \cite{Anisovich:2001pn}, the PWA shows that $a_4(2255)$ is a $^3F_4$ state.

In the reactions {$p\bar{p}\to \eta \pi^0\pi^0, \pi^0\pi^0, \pi^+\pi^-, \eta\eta, \eta\eta\prime$}, SPEC reported a ${^1G_4}$ state with the mass $2328\pm38 $ MeV and width $240\pm90$  MeV in the final states $(\pi a_2)_{L=4}$ and $(\pi a_0)_{L=4}$, where $L=4$ denotes $G$ wave \cite{Anisovich:2000ut}, which was named $\eta_4(2330)$. They also reported  the resonance $\pi_4(2250)$ in the $p\bar{p}$ annihilation through studying the Crystal Barrel data \cite{Anisovich:2001pn}.
The Regge trajectory analysis shows that both $\pi_4(2250)$ and $\eta_4(2330)$ are the ground states in the $\pi_4$ and $\eta_4$ families \cite{Anisovich:2002us,Anisovich:2003tm,Afonin:2007aa,Masjuan:2012gc} with different isospins $I=1$ and $0$, respectively. The study of mass spectrum of high-spin states in Ref. \cite {Ebert:2009ub} indicates that $\pi_4(2250)$ and $\eta_4(2330)$ are the first radial excitations of $G$-wave mesons, where a mass of 2092 MeV for the corresponding ground states was predicted.

$\omega_4(2250)$ \cite{Anisovich:2011sva} and  $\rho_4(2230)$ \cite{Anisovich:2002su} were reported by SPEC. Both the mass spectrum calculation in Ref. \cite {Ebert:2009ub} and the  $J-M^2$ plot in Ref. \cite{Afonin:2007aa,Masjuan:2012gc} support that $\omega_4(2250)$ and $\rho_4(2230)$ are the $1^3G_4$ states with different isospins, as above.

\subsection{States with $J=5$}

Analyzing the $\eta\omega$ and $\omega\pi^0\pi^0$ data, SPEC  strongly required a $^3G_5$ state around 2250 MeV \cite{Anisovich:2011sva}, which corresponds to $\omega_5(2250)$.
The mass error of $\omega_5(2250)$ was later given in Ref. \cite{Bugg:2004xu}.

An isospin $I=1$ and $J = 5$ structure $\rho_5(2350)$ was observed in the $p\bar{p}$  total cross section \cite{Abrams:1967zzb}, which can decay into  $\omega \pi^0, \pi^+ \pi^-, \pi^0\pi^0$, and $K^+K^-$ \cite{Alde:1994jm,Anisovich:2002su,Carter:1977jx,Hasan:1994he,Martin:1980ia,Alper:1980hi,Carter:1978ux}.

The present theoretical studies support $\omega_5(2250)$ as a ground state \cite{Masjuan:2012gc,Ebert:2009ub}.  $\rho_5(2350)$ is also a ground state, which was suggested in Refs.
\cite{ Anisovich:2000kxa,  Anisovich:2002us,   Anisovich:2003tm, Afonin:2007aa, Masjuan:2012gc, Ebert:2009ub}. The authors of Ref. \cite{Roca:2010tf}, however, treated $\rho_5(2350)$ as a molecule of four $\rho$ mesons.

\subsection{States with $J=6$}

GAM2  {observed }a $J=6$ neutral meson $R(2510)$ \cite{Binon:1983nx}; it is now named $f_6(2510)$ due to the contribution by Ref. \cite{Bolotov:1974iq}, where the branching ratio of its $\pi\pi$ mode was given. The $f_6(2510)$ was confirmed in the reaction
$\pi^-p\to 2\pi^0n$ \cite{Alde:1998mc} and SPEC also found it in the $p\bar{p}$ annihilation \cite{Anisovich:2000ut}.

There is only one experiment about $a_6(2450)$, which was observed by SPEC in the reaction $\pi p\to K_s^0K^{\pm}p$ \cite{Tseytlin:1988ne}.

From the mass spectrum analysis in Refs. \cite{Anisovich:2002us,Anisovich:2003tm,Ebert:2009ub}, the  $f_6(2510)$ is a good candidate of the $1^3H_6$ $q\bar{q}$ state. The $a_6(2450)$ is the isospin partner of $f_6(2510)$ \cite{ Anisovich:2000kxa,Anisovich:2002us,Anisovich:2003tm,Masjuan:2012gc,Ebert:2009ub}.
A different explanation for $f_6(2510)$, i.e., a molecular state of five $\rho$ mesons, was proposed in Ref. \cite{Roca:2010tf}.

From the above review, we can find that the present status of the high-spin states is still in disorder, where different groups gave different theoretical explanations.  This situation inspires us to carry out a systematic study of these high-spin states, in order to improve our understanding of the properties of these states.

\section{phenomenological analysis}\label{sec3}

The phenomenological analysis presented in this work includes two methodologies. First, the analysis of Regge trajectories is adopted to study possible meson assignments to the high-spin states under discussion. Second, we use the QPC model to obtain their OZI-allowed two-body decay behaviors. In the following, we give a brief introduction to these methods.

The analysis of Regge trajectory provides a general method to study the meson spectrum \cite{Anisovich:2000kxa}. The excited states and ground states satisfy a simple relation
\begin{equation}
M^{2}=M_{0}^{2}+\mu^{2}(n-1),\label{nm}
\end{equation}
where $M_{0}$ and $M$ are the masses of ground state and excited state, respectively. The $\mu^{2}$ gives a slope of a trajectory with the value $\mu^{2}=1.25\pm0.15$ GeV$^{2}$ suggested in Ref. \cite{Anisovich:2000kxa}. Via the above equation, we obtain the $n$-$M^2$ plot of the mesons under discussion, where the radial quantum number $n$ of these states can be obtained when mass is given, this is important information regarding the underlying structure of mesons.

In addition to the relation in Eq. (\ref{nm}), there exists a similar relation
\begin{equation}
M_{J}^{2}=M_{J^\prime}^{2}+\alpha^{2}(J-J^\prime),\label{jm}
\end{equation}
where $J$ or $J^\prime$ denotes the spin of a meson. $M_{J^\prime}$ and $M_{J}$ are the masses of mesons with different spins and with the same $P$ and $C$ quantum numbers. Via Eq. (\ref{jm}), the corresponding $J$-$M_J^2$ plot can be obtained, which provides an extra test of the conclusion from the $n$-$M^2$ plot. 

{When further checking the relation of masses of high-spin mesons with the principle quantum number $N=n+J$, 
we find that there exists a symmetry of the spectrum in the form
\begin{equation}
M_{N}^{2}=M_{0}^{2}+\beta^{2}N,\label{Nm}
\end{equation}
where this phenomenon argues in favor of the existence of the principle quantum that governs the spectrum of excited mesons  indicated in Refs. \cite{Glozman:2007at, Afonin:2006wt}.}

In the following, we briefly explain the QPC model adopted in this work. After the
QPC model was proposed by Micu \cite{Micu:1968mk}, it was further developed by the Orsay group  
\cite{Le Yaouanc:1972ae,Le Yaouanc:1973xz,Le Yaouanc:1974mr,Le Yaouanc:1977ux,Le Yaouanc:1977gm}.
 Later, the model was widely applied  to study the OZI-allowed strong decay of hadrons
 \cite{vanBeveren:1982qb,Capstick:1993kb,Blundell:1995ev,Ackleh:1996yt,Capstick:1996ib,Close:2005se,Zhang:2006yj,Lu:2006ry,Sun:2009tg,Liu:2009fe,
Sun:2010pg,Rijken:2010zza,Yu:2011ta,Zhou:2011sp,Ye:2012gu,Sun:2013qca,He:2013ttg,Sun:2014wea,Pang:2014laa,Wang:2014sea,Wang:2012wa,Chen:2015iqa}.

For a two-body strong decay process $A\rightarrow B+C$, the corresponding transition matrix element can be written as
\begin{eqnarray}
\langle BC|T|A \rangle = \delta ^3(\mathbf{P}_B+\mathbf{P}_C)\mathcal{M}^{{M}_{J_{A}}M_{J_{B}}M_{J_{C}}},
\end{eqnarray}
where $\mathbf{P}_{B(C)}$ denotes the three-momentum of a final particle $B(C)$. $M_{J_i}$ ($i=A,\,B,\,C$) is an orbital magnetic momentum of the corresponding meson in the decay. $\mathcal{M}^{{M}_{J_{A}}M_{J_{B}}M_{J_{C}}}$ is the amplitude we calculate. The $T$ operator reads as
\begin{eqnarray}
T& = &-3\gamma \sum_{m}\langle1m;1~-m|00\rangle\int d \mathbf{p}_3d\mathbf{p}_4\delta ^3 (\mathbf{p}_3+\mathbf{p}_4) \nonumber \\
 && \times \mathcal{Y}_{1m}\left(\frac{\textbf{p}_3-\mathbf{p}_4}{2}\right)\chi _{1,-m}^{34}\phi _{0}^{34}
\left(\omega_{0}^{34}\right)_{ij}b_{3i}^{\dag}(\mathbf{p}_3)d_{4j}^{\dag}(\mathbf{p}_4),
\end{eqnarray}
where $\gamma$ is a parameter that takes the value 8.7 or ${8.7}/{\sqrt{3}}$ when the quark-antiquark pair created from the vacuum
is $u\bar u(d\bar d)$ or $s\bar s$ \cite{Ye:2012gu}. The quark and antiquark created from the vacuum are marked by the subscripts 3 and 4, respectively. $i/j$ denotes the color indexes, $\chi$, $\phi$, and $\omega$  are the spin, flavor and color wave functions, respectively, and $\mathcal{Y}_{\ell m}(\mathbf{p})=|\mathbf{p}|^\ell Y_{lm}(\mathbf{p}) $ is the  solid harmonic polynomial
(see Refs. \cite{Roberts:1992js,Blundell:1996as} for more details). Using the Jacob-Wick formula \cite{Jacob:1959at}, the amplitude $\mathcal{M}^{{M}_{J_{A}}M_{J_{B}}M_{J_{C}}}$
can be converted into the partial wave amplitude $M^{JL}(\mathbf{P})$, i.e.,
\begin{eqnarray}
\mathcal{M}^{JL}(\mathbf{P})&=&\frac{\sqrt{4\pi(2L+1)}}{2J_A+1}\sum_{M_{J_B}M_{J_C}}\langle L0;JM_{J_A}|J_AM_{J_A}\rangle \nonumber \\
&&\times \langle J_BM_{J_B};J_CM_{J_C}|{J_A}M_{J_A}\rangle \mathcal{M}^{M_{J_{A}}M_{J_B}M_{J_C}}.
\end{eqnarray}
Finally, the decay width can be given by
\begin{eqnarray}
\Gamma&=&\frac{\pi|\mathbf{P}|}{4m_A^2}\sum_{J,L}|\mathcal{M}^{JL}(\mathbf{P})|^2,
\end{eqnarray}
where $m_A$ is the mass of the initial meson $A$.
In the concrete calculation, we use the harmonic oscillator wave function to describe the
meson spatial wave function. The harmonic oscillator wave function has the following expression
$\Psi_{nlm}(R,\mathbf{p})= \mathcal{R}_{nl}(R,\mathbf{p}) \mathcal{Y}_{lm}(\mathbf{p})$,
with $R$ being a parameter, which is given in Ref.~\cite{Close:2005se} for the mesons involved in our calculation.

Before performing the phenomenological analysis of these high-spin mesons, we need to emphasize that 
the orbital quantum number for the high-spin mesons with  $J^{PC}=3^{--}$, $4^{++}$, $5^{--}$, and $6^{++}$
cannot be fixed \cite{Glozman:2007at, Glozman:2007ek,Glozman:2014mka}. 
For example, the meson with $J^{PC}=3^{--}$ is the mixture between the $L=2$ and $L=4$ states\footnote{The meson with $J^{PC}=4^{++}$ is the mixture between the $L=3$ and $L=5$ states, the meson with $J^{PC}=5^{--}$ is from the mixture between the $L=4$ and $L=6$ states, and the meson with $J^{PC}=6^{++}$ is due to the mixture between the $L=5$ and $L=7$ states}. By our calculation, we find that the contribution of the higher orbital quantum number for these high-spin mesons with $J^{PC}=3^{--}$, $4^{++}$, $5^{--}$, and $6^{++}$
is far smaller than that of  the lower orbital quantum number when we study their decay behavior. Thus, in the following, we will only consider contributions from the lower orbital quantum number for the discussed high-spin mesons with  $J^{PC}=3^{--}$, $4^{++}$, $5^{--}$, and $6^{++}$. 

\subsection{Twelve $J^{--}$ states}

In this subsection, we discuss twelve observed high-spin states with  $J^{--}$ ($J=3,4,5$) quantum numbers (see Table \ref{tab:1}).
The corresponding analysis of Regge trajectories with the $n$-$M^2$ and $J$-$M^2$ plots are shown in Fig. \ref{omega3njm}.

There are eight high-spin states with $J^{PC}=3^{--}$, which are $\omega_3(1670)$, $\omega_3(1945)$, $\omega_3(2255)$, $\omega_3(2285)$, $\phi_3(1850)$, $\rho_3(1690)$, $\rho_3(1990)$, and $\rho_3(2250)$.  As it is an effective approach to study the meson categorization, analysis of the Regge trajectory is applied to further discuss  $\omega_3(1670)$, $\omega_3(1945)$, $\omega_3(2255)$, $\omega_3(2285)$, and $\phi_3(1850)$. In Ref. \cite{He:2013ttg}, the properties of $\rho_3(1690)$, $\rho_3(1990)$, and $\rho_3(2250)$ were studied, where they can be explained as $n^3D_3$ ($n=1,2,3$) states in the $\rho_3$ meson family, respectively.
In Fig. \ref{omega3njm} (a), we make a comparison of the observed $\omega_3$ and $\rho_3$ states, which reflects the similarity between the $\rho_3$ and $\omega_3$ meson families that is due to their similar values of the slope $\mu^2$.
Thus, we can conclude that $\omega_3(1670)$, $\omega_3(1945)$, and $\omega_3(2285)$ are the isospin partners of $\rho_3(1690)$, $\rho_3(1990)$, and $\rho_3(2250)$, respectively.
\begin{figure}[htbp]
\begin{center}
\scalebox{1}{\includegraphics[width=\columnwidth]{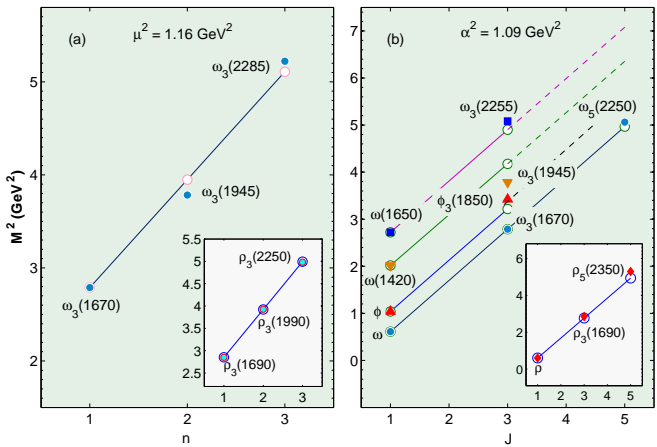}}
\caption{(color online). Analysis of Regge trajectories for the observed states with  $J^{--}$ quantum numbers. Diagram (a) is he $n$-$M^2$ plots for $\omega_3$ and $\rho_3$ states, while diagram (b) is the corresponding $J$-$M^2$ plots for $\omega$, $\phi$, and $\rho$. {Here, open and filled circles are the theoretical and experimental values, respectively, and  $\rho(770)$, $\omega(782)$, and $\phi(1020)$ are abbreviated as $\rho$, $\omega$, and $\phi$, respectively; these conventions are adopted in the following figures.}} \label{omega3njm}
\end{center}
\end{figure}

As shown in Fig. \ref{omega3njm} (b), we also give the $J$-$M^2$ plot analysis, which also supports the assignments of $\omega_3(1670)$, $\omega_3(1945)$,  and $\omega_3(2285)$ as the ground state, first, and second radial excitations, respectively. In addition, the $J$-$M^2$ analysis also indicates that $\phi_3(1850)$ is the ground state in the $\phi_3$ meson family.

The partial wave analysis in Ref. \cite{Anisovich:2011sva} indicates that $\omega_3(2255)$ is a $G$-wave meson. Furthermore, $\omega_3(2255)$ corresponds to $\omega_3(1^3G_3)$, an assignment that is supported by the $J$-$M^2$ plot  in Fig. \ref{omega3njm} (b). Later, we will discuss the decay behavior of $\omega_3(2255)$ corresponding to this assignment.

We notice that the mass $\omega_4(2250)$ is close to that of $\rho_4(2230)$, which shows that it is reasonable to assign $\omega_4(2250)$ as the isospin partner of $\rho_4(2230)$. In this work, $\omega_4(2250)$ and $\rho_4(2230)$ are treated as  $\omega_4(1^3G_4)$ and $\rho_4(1^3G_4)$, respectively.

According to the $J$-$M^2$ plot shown in Fig. \ref{omega3njm} (b), we can conclude that $\omega_5(2250)$ and  $\rho_5(2350)$ are mesons with quantum number $1^3G_5$.

In the following, we further present the study of their two-body OZI-allowed decays.

\begin{figure*}[htbp]
\begin{center}
\scalebox{2}{\includegraphics[width=\columnwidth]{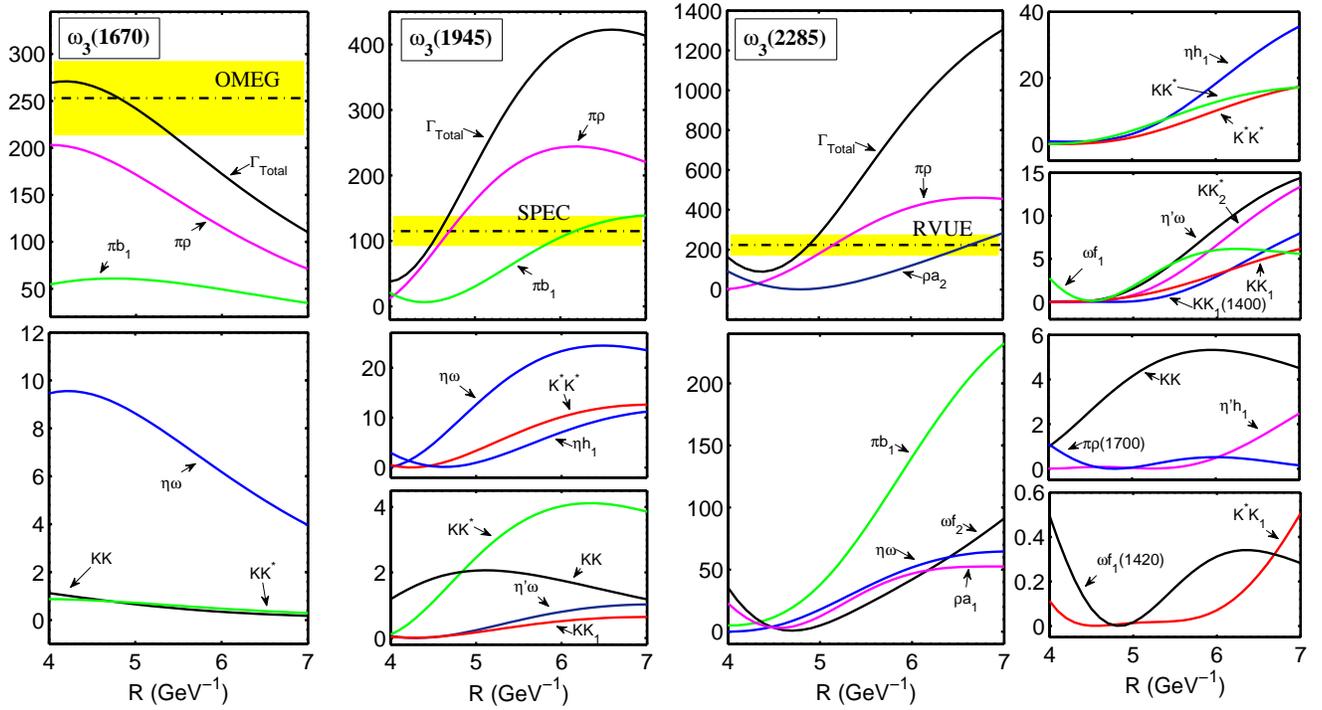}}
\caption{(color online).  The $R$-value dependence of the partial and total decay widths of $\omega_3(1670)$, $\omega_3(1945)$, and $\omega_3(2285)$ states.  All results are in units of MeV. The dot-dashed lines with yellow bands are the corresponding experimental widths of $\omega_3(1670)$ \cite{Corden:1977xu}, $\omega_3(1945)$ \cite{Anisovich:2011sva}, and $\omega_3(2285)$ \cite{Anisovich:2011sva}.  {Here, $a_1(1260)$, $h_1(1170)$, $f_1(1285)$, $K^*(892)$, $K_1(1270)$, and $K_2^*(1430)$ are abbreviated as $a_1$, $h_1$, $f_1$ , $K^*$, $K_1$, and  $K_2^*$, respectively; these conventions  are  adopted in the following figures.}} \label{fig:omega3}
\end{center}
\end{figure*}

\begin{figure*}[htbp]
\begin{center}
\scalebox{1.6}{\includegraphics[width=\columnwidth]{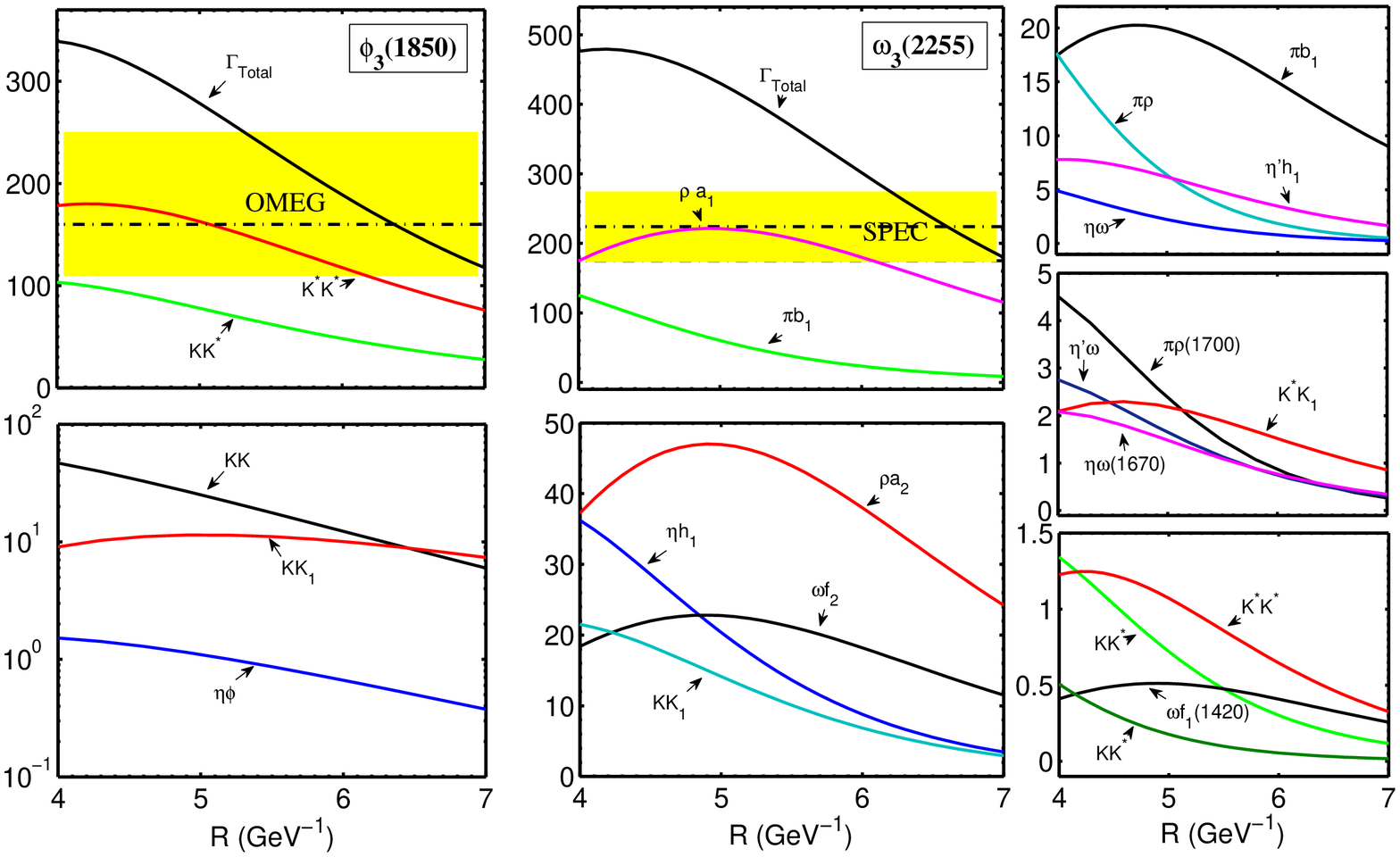}}
\caption{(color online).  The $R$-value dependence of the partial and total decay widths of the $\omega_3(2255)$ and $\phi_3(1850)$ states under discussion.  All results are in units of MeV. The dot-dashed lines with yellow bands are the corresponding experimental widths of $\omega_3(2255)$ \cite{Anisovich:2011sva} and $\phi_3(1850)$ \cite{Armstrong:1982jj}. } \label{fig:omega3phi3}
\end{center}
\end{figure*}

\subsubsection{$\omega_3(1670)$, $\omega_3(1945)$, $\omega_3(2255)$, $\omega_3(2285)$, and $\phi_3(1850)$}

As for $\omega_3(1670)$, its two-body OZI-allowed strong decays with a $1^3D_3$ assignment are given in Fig. \ref{fig:omega3}. When taking $R  = 4.0 - 5.4$ GeV$^{-1}$, our theoretical result overlaps with the experimental width of $\omega_3(1670)$ \cite{Corden:1977xu}\footnote{Just shown in PDG \cite{Agashe:2014kda}, different experiments gave different results of the width of $\omega_3(1670)$. When comparing our calculation with experimental data, we adopt the result in Ref. \cite{Corden:1977xu} since the corresponding $R$ value is reasonable. In the following discussion, we
notice the adopted $R$ ranges for $\omega_3(1945)$ and $ \omega_3(2285)$, which satisfy the requirement that the $R$ range becomes more larger with increasing the radial quantum number.}. Here, $\pi\rho$ and $\pi b_1$ are main decay modes of $\omega_3(1670)$; this is consistent with the experimental observation because $\pi\rho$ and $\pi\omega\pi$ channels were reported in experiment and $b_1(1235)$ dominantly decays into $\pi\omega$.  In summary, $\omega_3(1670)$ as $\omega_3(1^3D_3)$ meson is possible: this was addressed in Refs. \cite{Bugg:2004rj,Klempt:2007cp,Anisovich:2002us,Masjuan:2012gc}.

We present the decay behavior of $\omega_3(1945)$  under the $2^3D_3$ state assignment, in Fig. \ref{fig:omega3}. The obtained total decay width can reproduce the experimental width of $\omega_3(1945)$ measured in Ref. \cite{Anisovich:2011sva} if $R=4.5-4.7$ GeV$^{-1}$. Our results also show that $\pi\rho$, $\pi b_1$, and $\eta\omega$ are its main decay channels. As  $\omega_3(1945)\to \eta\omega$ was reported in Ref. \cite{Anisovich:2011sva}, we thus  suggest further experimental searches for  decay channels $\pi\rho$ and $\pi b_1$, which would be useful to test the underlying  structure of $\omega_3(1945)$.

The decay properties of $\omega_3(2285)$ are given in Fig. \ref{fig:omega3}. When taking $R=4.7-5.0$ GeV$^{ -1}$, our theoretical results are consistent with the measured experimental width \cite{Anisovich:2011sva}. In addition, $\omega_3(2285)$ mainly decays into {$\pi\rho$,  $\pi b_1$,  and $\eta\omega$}, which could explain why $\omega_3(2285)$ was first observed in the $\eta\omega$ channel \cite{Anisovich:2011sva}. Experimental exploration of $\omega_3(2285)$ via the several remaining  main decay modes predicted in this work would be an  an interesting area for investigation.

The above studies indicate that description of $\omega_3(1670)$, $\omega_3(1945)$, and $\omega_3(2285)$ as the ground state, first and  second radial excitations, respectively, should be further tested.

Next, we illustrate the decay behaviors of $\omega_3(2255)$ under the $\omega_3(1^3G_3)$ assignment (see Fig. \ref{fig:omega3phi3} for the details). Its dominant decay channels are $\rho a_1$, $\pi b_1$,  $\eta \omega $, and $\rho a_2$. Additionally, the channels $\omega f_2$, and $\eta h_1$ also contribute, mainly  to the total width. These quantitative predictions can  serve as further experimental investigation of $\omega_3(2255)$. We need to specify that the obtained total decay width is strongly dependent on the range of $R$ value. Experimental data of the width of $\omega_3(2255)$ was given by Ref. \cite{Anisovich:2011sva}, which can be reproduced by our calculation with $R\sim 7$ GeV$^{-1}$.
For the experimental study of $\omega_3(2255)$, a crucial task is the precise measurement of its resonance parameter, which can provide  more abundant information for identifying  $\omega_3(2255)$ as the $1^3G_3$ assignment.

There is only one $\phi_3$ state listed in the  PDG, i.e., $\phi_3(1850)$. We calculate its two-body decays under the $\phi_3(1^3D_3)$ assignment, listed in Fig. \ref{fig:omega3phi3}. Here, $K^*K^*$, $KK^*$, $KK$, and $KK_1(1270)$ are the main decay modes of $\phi_3(1850)$. We notice that experimental data of ratio ${\Gamma_{K K^*}}/{\Gamma_{KK}}=0.55^{+0.85}_{ - 0.45}$ \cite{Aston:1988rf} shows that the partial width of  the $KK$ mode is larger than that of the $KK^*$ mode. However, our result shows that
the partial width of the $KK$ mode is smaller than that of the $KK^*$ mode
for $\phi_3(1850)$,  since we obtain ${\Gamma_{K K^*}}/{\Gamma_{KK}}=3.5 - 18$. Our conclusion of the $KK$ and $KK^*$ channels is also supported by the study presented in Ref. \cite{Bramon:1980xm}, where the authors obtained $\Gamma_{KK}=43\pm4 $ MeV and  $\Gamma_{KK^*}=55\pm 10$ MeV.
Because there is only one experimental measurement for this ratio at present, we expect future experiments to clarify the above inconsistency between  theoretical and experimental results.
Additionally, when we  take $R = 5.3-7.0$ GeV$^{-1}$, we can find a common range where the theoretical  result is consistent with the experimental width given in Ref. \cite{Armstrong:1982jj}.

\subsubsection{$\omega_4(2250)$, $\rho_4(2230)$, $\omega_5(2250)$ and $\rho_5(2350)$}

\begin{figure*}[htbp]
\begin{center}
\scalebox{1.95}{\includegraphics[width=\columnwidth]{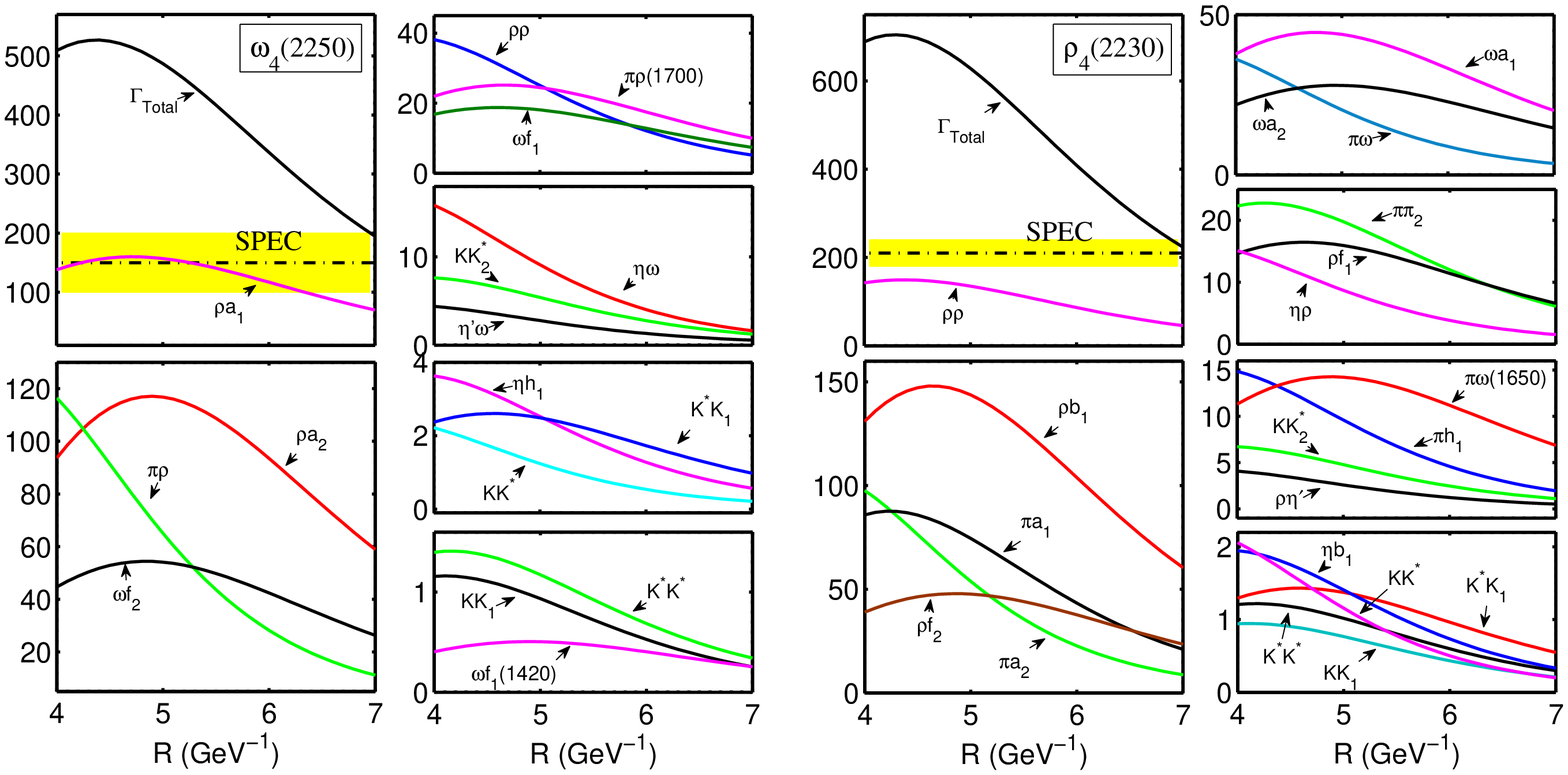}}
\caption{(color online). The $R$-value dependence of the partial and total decay widths of $\omega_4(2250)$ and $\rho_4(2230)$ states.  All results are in units of MeV. The dot-dashed lines with yellow bands denote the corresponding experimental widths of  $\omega_4(2250)$ \cite{Anisovich:2011sva} and $\rho_4(2230)$ \cite{Anisovich:2002su}. {Here, $\pi_2(1670)$ is abbreviated as $\pi_2$}.} \label{fig:omega4rho4}
\end{center}
\end{figure*}

As the candidates of $\omega_4(1^3G_4)$ and $\rho_4(1^3G_4)$, respectively, 
$\omega_4(2250)$ and  $\rho_4(2230)$ have the decay behaviors listed in Fig. \ref{fig:omega4rho4}, we conclude, with this study, that $\rho a_1$, $\rho a_2$, and $\pi \rho$ are the main decay modes of $\omega_4(2250)$, while $\rho_4(2230)$ mainly decays into $\rho\rho$, $\rho b_1$, $\pi a_2$, and $\pi a_1$. At present, experimental information for  $\omega_4(2250)$ and  $\rho_4(2230)$ is still scarce. For example, there is only one experimental measurement for the widths of $\omega_4(2250)$ and $\rho_4(2230)$.
When $R\approx 7$ GeV$^{ - 1}$ is adopted, theoretical total decay widths can  overlap with the experimental data  for $\omega_4(2250)$ \cite{Anisovich:2011sva} and $\rho_4(2230)$ \cite{Anisovich:2002su}.

\begin{figure*}[htbp]
\begin{center}
\scalebox{1.95}{\includegraphics[width=\columnwidth]{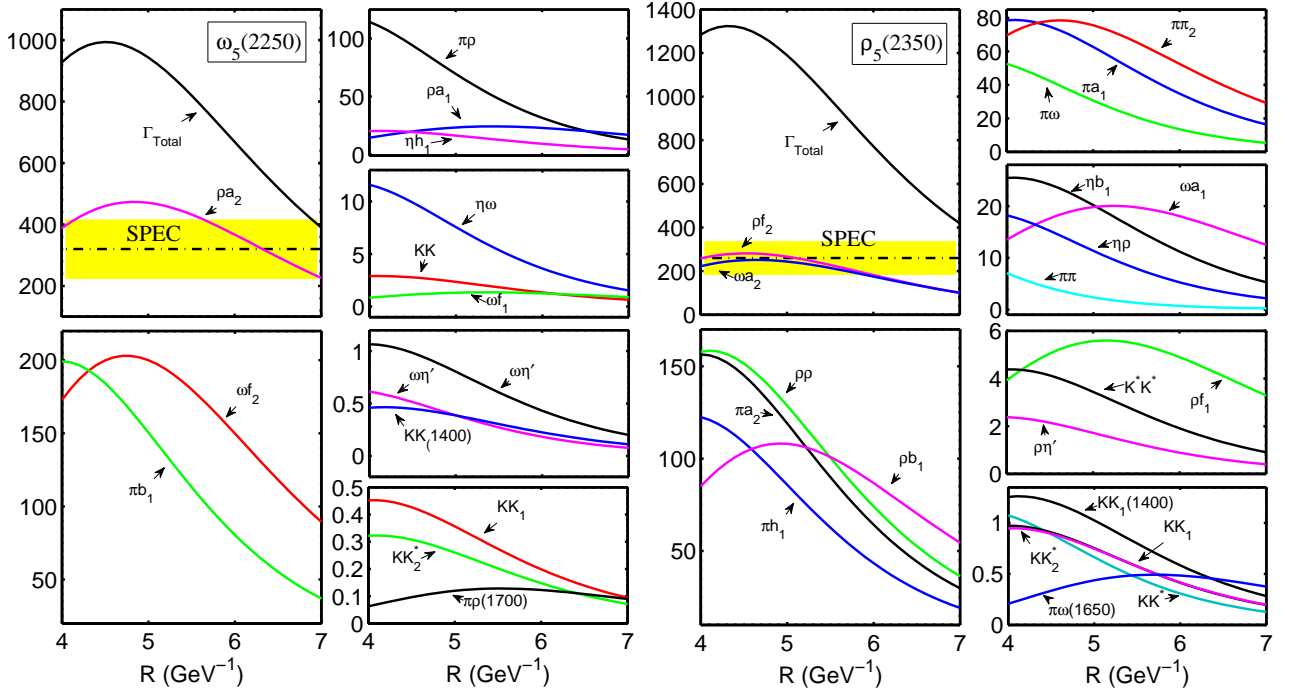}}
\caption{(color online).  The $R$-value dependence of the partial and total decay widths of $\omega_5(2250)$ and $\rho_5(2350)$ states.  All results are in units of MeV. The dot-dashed lines with yellow bands are experimental widths of $\omega_5(2250)$ \cite{Bugg:2003kj} and $\rho_5(2350)$ \cite{Alde:1994jm}.} \label{fig:omega5rho5}
\end{center}
\end{figure*}

Both $\omega_5(2250)$ and $\rho_5(2350)$ are good candidates for $G$-wave mesons. In Fig. \ref{fig:omega5rho5}, we present their decay features. Here, the main decay modes of $\omega_5(2250)$ include
$\rho a_2$, $\omega f_2$,  $\pi b_1$ and $\pi \rho$, among which
$\pi b_1$ was reported in experiment \cite{Bugg:2003kj}. The dominant decay modes of $\rho_5(2350)$ are 
$\rho f _2$ and $\omega a_2$, while  $\rho\rho$, $\pi a_2$ and $\pi h_1$ are also important contributions to the total decay width. Because the experimental status of $\omega_5(2250)$ and $\rho_5(2350)$ is similar to that of $\omega_4(2250)$ and  $\rho_4(2230)$, there is not enough information on their experimental data. Thus, we compare the obtained total widths of $\omega_5(2250)$ and $\rho_5(2350)$ with the present experimental data \cite{Bugg:2003kj,Alde:1994jm} (see Fig. \ref{fig:omega5rho5} for more details).

These theoretical predictions of the decay behaviors of $\omega_4(2250)$, $\rho_4(2230)$, $\omega_5(2250)$ and $\rho_5(2350)$ will be useful for future experimental studies.

\subsection{Eleven $J^{++}$ states}

As listed in Table \ref{tab:1}, 11 high-spin states with the $J^{++}$ ($J=3,4,6$) quantum numbers were reported in experiments. 
In Fig. \ref{fanjm}, we present the systematic analysis of Regge trajectories with the $n$-$M^2$ and $J$-$M^2$ plots, which is helpful for obtaining the information regarding their classification into meson families.

\begin{figure}[htbp]
\begin{center}
\scalebox{1}{\includegraphics[width=\columnwidth]{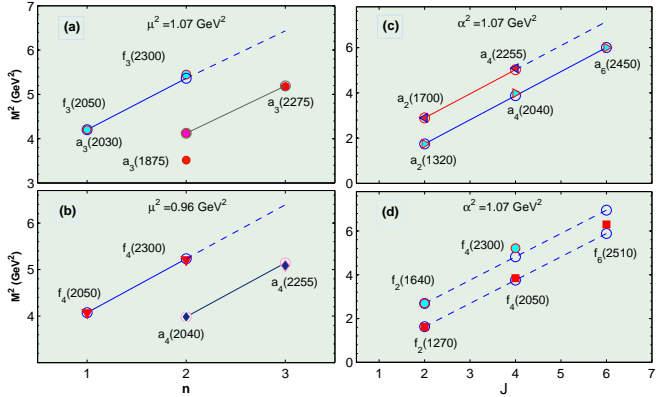}}
\caption{(color online). The analysis of Regge trajectories for the observed states with  $J^{++}$ quantum numbers.  (a) and (b) are the $n$-$M^2$ plots for the $3^{++}$ and $4^{++}$ states, respectively, while  (c) and (d) are the corresponding $J$-$M^2$ plots. For  (a) and (b), the $I=1$ $n$-$M^2$ trajectories are
displaced one unit to the right in $n$ in order to resolve them from $I=0$.} \label{fanjm}
\end{center}
\end{figure}

Figure. \ref{fanjm} (a) shows that $f_3(2050)$ is the ground state of  the $f_3$ family, while $f_3(2300)$ is the radial excitation, which is in good agreement with the conclusion in Refs. \cite{Anisovich:2000ut,Afonin:2007aa}. 
As the isospin parters of $f_3(2050)$ and $f_3(2300)$, $a_3(2030)$ and $a_3(2275)$
are the ground state and the first radial excitation in the {$a_3$} meson family, respectively, which is reflected in Fig. \ref{fanjm} (a). 
In addition, we notice the $a_3(1875)$ state, which cannot be categorized into the {$a_3$}  meson family. 
Thus, in the following discussion of their decay behaviors, we will mainly focus on $f_3(2050)$, $f_3(2300)$, $a_3(2030)$, and $a_3(2275)$. 

When checking $a_4(2040)$ and $a_4(2255)$ and  their isospin partners $f_4(2050)$ and $f_4(2300)$, we conclude that 
$a_4(2040)/f_4(2050)$ and $a_4(2255)/f_4(2300)$ are the ground state and the first radial excitation in the $a_4/f_4$ meson families, respectively; this can be supported by the analysis of the $n$-$M^2$ and $J$-$M^2$ plots shown in Fig. \ref{fanjm}.   

From the analysis shown in Fig. \ref{fanjm}, we can  conclude that $f_6(2510)$ is a candidate of the $f_6(1^3H_6)$ meson and  that $a_6(2450)$ is the isospin partner of $f_6(2510)$.

In the following, we further test the above-mentioned meson assignments to the states with $J^{++}$ quantum numbers by studying  their two-body OZI-allowed strong decays. 

\subsubsection{$f_3(2050)$, $f_3(2300)$, $a_3(2030)$, and $a_3(2275)$}

 \begin{figure*}[htbp]
\begin{center}
\scalebox{1.9}{\includegraphics[width=\columnwidth]{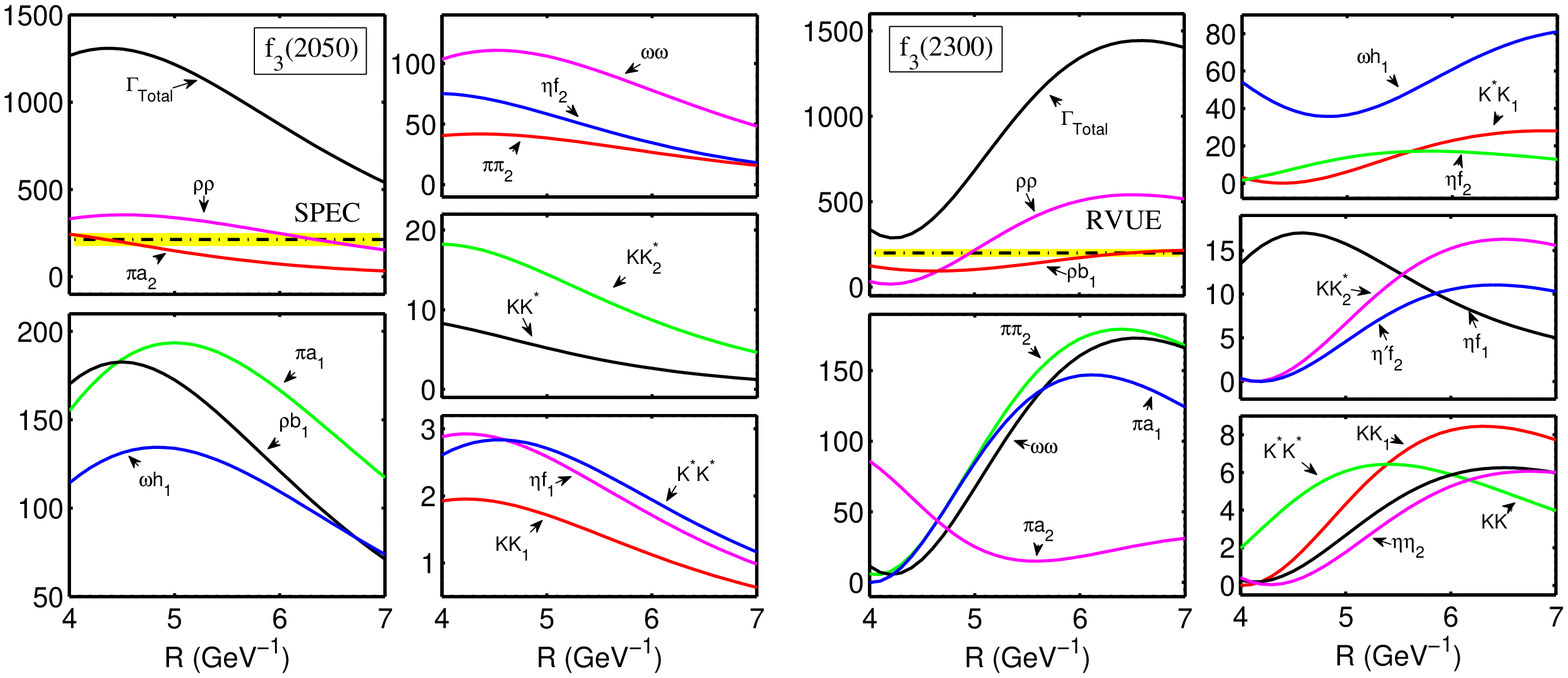}}
\caption{(color online).  The $R$ dependence of the partial and total decay widths of $f_3(2050)$ and $f_3(2300)$ states.  All results are in units of MeV. The dot-dashed lines with yellow bands are the corresponding experimental widths of $f_3(2050)$ \cite{Anisovich:2000ut}, and $f_3(2300)$ \cite{Bugg:2004rj}. {Here, $\eta_2(1645)$ is abbreviated as $\eta_2$, as adopted though this work.}} \label{f3}
\end{center}
\end{figure*}

 \begin{figure*}[htbp]
\begin{center}
\scalebox{1.9}{\includegraphics[width=\columnwidth]{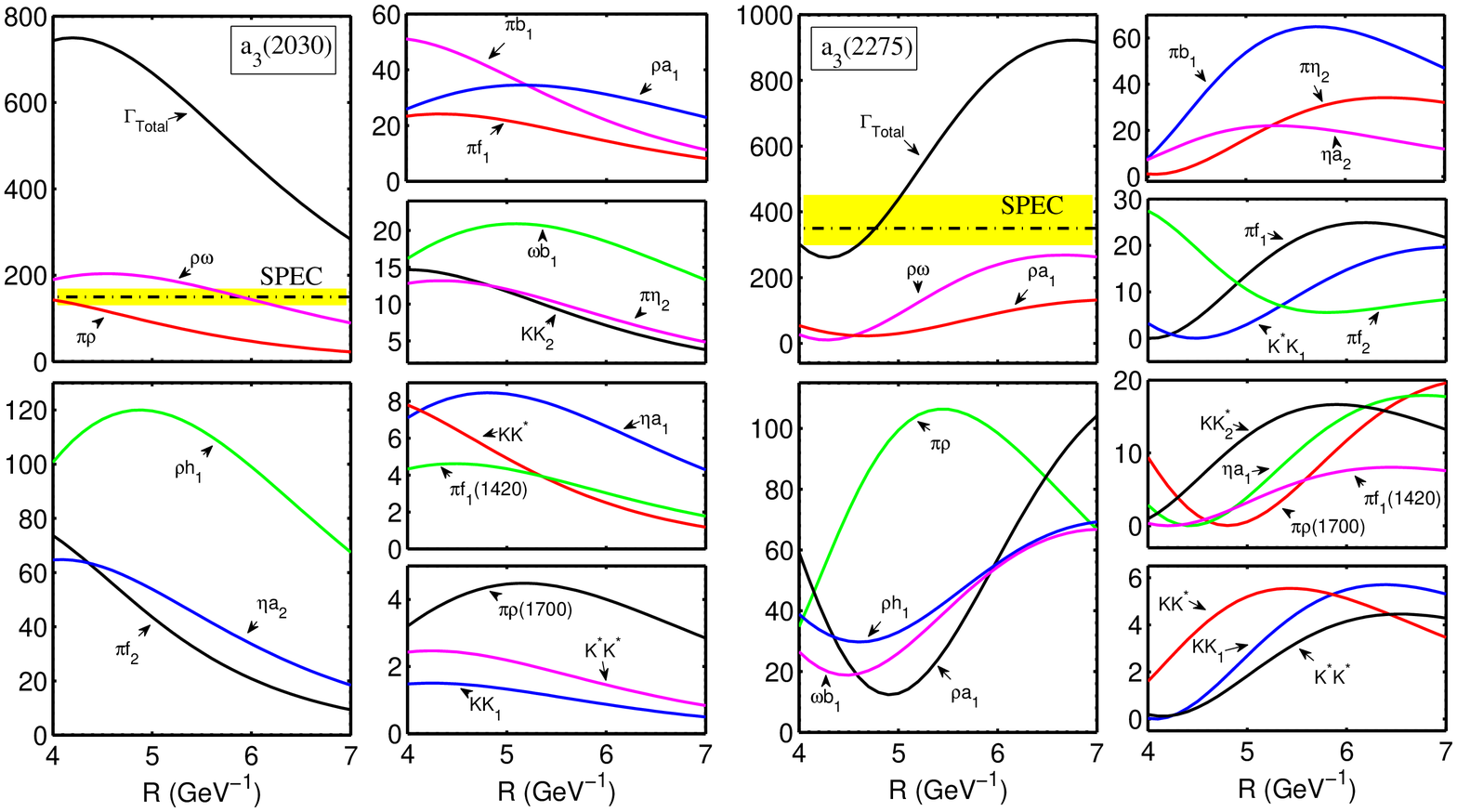}}
\caption{(color online).  The $R$ dependence of the partial and total decay widths of $a_3(2030)$ and $a_3(2275)$ states.  All results are in units of MeV. The dot-dashed lines with yellow bands are the corresponding experimental widths of  $a_3(2030)$ \cite{Anisovich:2001pn}, and $a_3(2275)$ \cite{Anisovich:2001pp}. {Here, $K_2(1580)$ is abbreviated as  $K_2$ adopted in this work.}} \label{a3}
\end{center}
\end{figure*}

In Figs \ref{f3}-\ref{f4}, the decay behaviors of $f_3(2050)$, $f_3(2300)$, $a_3(2030)$, and $a_3(2275)$ are given, where their partial and total decay widths are obtained by the QPC model. We also compare our results with the experimental data. 

$f_3(2050)$ and $f_3(2300)$ are treated as $f_3(1^3F_3)$ and $f_3(2^3F^3)$,  respectively, with isospin $I=0$. 
 Our calculation  for $f_3(2050)$ indicates that $\rho\rho$, $\pi a_2$, $\rho b_1$, and $\pi a_1$ are its main decay channels, and 
$f_3(2050)\to \eta f_2$ is a sizable contribution to its total decay width, this could explain why  experiments reported $f_3(2050)$ in its $\pi a_2$ and $\eta f_2$ decay modes. However, the obtained total decay width of $f_3(2050)$ is far larger than the experimental width given in Ref. \cite{Anisovich:2000ut} when taking $R=4 - 7$ GeV$^{-1}$. 
For $f_3(2300)$, the main decay channels include $\rho\rho$, $\rho b_1$, $\pi \pi_2$, and $\omega\omega$. 
The calculated total decay width of $f_3(2300)$ is also larger than the experimental data \cite{Anisovich:2000ut} (see Fig. \ref{f3}). 
The situation of $f_3(2050)$ is similar, to an extent, to that of $f_3(2050)$. To further clarify the above inconsistency between the experimental width and the theoretical result, further experimental measurement of $f_3(2050)$ and $f_3(2300)$ is encouraged. 

Before illustrating the decay properties of $a_3(2030)$, and $a_3(2275)$, the isospin partners  of $f_3(2050)$ and $f_3(2300)$, 
we still need to discuss $a_3(1875)$. Although $a_3(1875)$ cannot be grouped into the $a_3$ meson family  in term of the analysis of the Regge trajectories only, the authors of Ref. \cite{Masjuan:2012gc} suggest that $a_3(1875)$ and $a_3(2030)$ are the same state. If $a_3(1875)$ is $a_3(1^3F_3)$, our calculated results of the branching ratio 
 $B(a_3(1875)\to f_2(1270)\pi)/B(a_3(1875)\to \pi\rho)$ is about $1$, which is consistent with the experimental data $0.8\pm 0.2$ given in Ref. \cite{Chung:2002pu}. Additionally, $B(a_3(1875)\to \rho_3(1690)\pi)/B(a_3(1875)\to \pi\rho)$ in Ref. \cite{Chung:2002pu} is about $0.9\pm0.3$, where our calculation gives $1.9 - 2.4$ for this ratio. The $a_3(1^3F_3)$ assignment to $a_3(1875)$, therefor, seems to be reasonable. Thus, measuring the resonance parameters of
$a_3(1875)$ and $a_3(2030)$ is  crucial to test whether $a_3(1875)$ and $a_3(2030)$ are the same state. 

$a_3(2030)$ mainly decays into $\rho\omega$, $\pi\rho$, and $\rho h_1$, and $\eta a_2$ and $\pi f_2$ sizably contribute to the total width. 
Here, the $a_3(2030)$ decays into $\eta a_2$ and $\pi f_2$ were observed in experiment \cite{Agashe:2014kda}. 
The theoretical total decay width of $a_3(2030)$ with the range $R=4 - 7$ GeV$^{-1}$ is far larger than the experimental measurement \cite{Anisovich:2001pn}. 
Under the $a_3(1^3F_3)$ meson assignment, $a_3(2275)$ has main decay modes $\rho\omega$, $\rho a_1$, and $\pi\rho$. 
Figure. \ref{a3} displays the $R$ dependence of the partial decay width of $a_3(2275)$; this shows that the calculated total width overlaps with the experimental data \cite{Anisovich:2001pp} when $R=4.6 - 5$ GeV$^{ - 1}$.

\subsubsection{$f_4(2050)$, $f_4(2300)$, {$a_4(2040)$}, and $a_4(2255)$}

In this subsection, we present the decay properties for four $4^{++}$ states $f_4(2050)$, $f_4(2300)$, {$a_4(2040)$}, and $a_4(2255)$, which are shown in Figs.  \ref{f4}-\ref{a4}.

\begin{table}[htbp]
\caption{Comparison between the calculated and experimental results for some typical ratios of $f_4(2050)$. Here, the theoretical results  are obtained by taking $R=4.0 - 7.0$ GeV$^{-1}$.}  \label{tab:f4}
  \renewcommand{\arraystretch}{1.5}
 \tabcolsep=2pt
\[\begin{array}{ccccccc}
\toprule[1pt]
{\rm Ratios}&{\rm This\,work}&{\rm Experiment}\\
\midrule[1pt]
{\Gamma_{KK}}/{\Gamma_{\pi\pi}}&0.019-0.025&0.04^{+0.02}_{ - 0.01}$ \cite{Etkin:1981sg}$ \\
{\Gamma_{\pi\pi}}/{\Gamma_{Total}} &0.006 - 0.036&0.170\pm0.015 $ \cite{Agashe:2014kda}$\\
{\Gamma_{\omega\omega}}/{\Gamma_{\pi\pi}}&3.9-21&1.5\pm0.3 $ \cite{Etkin:1981sg}$ \\
{\Gamma_{\eta\eta}}/{\Gamma_{Total}}&(0.25-1.3)\times 10^{-3}&(2.1\pm0.8)\times10^{ - 3}  $ \cite{Alde:1985kp}$\\
\bottomrule[1pt]
\end{array}\]
\end{table}

The results of $f_4(2050)$ shown in Fig.  \ref{f4} indicate that $\rho\rho$, $\pi a_2$, and $\omega\omega$ are its dominant  decay channels. Furthermore, we also obtain some typical ratios, which are comparable with the experimental data (see Table \ref{tab:f4} for more details).  For $f_4(2300)$, $\pi a_2$ and $\rho\rho$ are its dominant  decay channels, and the obtained ratio ${\Gamma_{f_4(2300)\to \rho\rho}}/{\Gamma_{f_4(2300)\to \omega\omega}} = 0.6  -  3.1$ is consistent with the experimental value $2.8\pm 0.5$ \cite{Barberis:2000kc}. The total decay widths of $f_4(2050)$ and $f_4(2300)$ overlap with the corresponding experimental widths when $R=4 -  7$ GeV$^{-1}$. 
Thus, these studies support $f_4(2050)$ and $f_4(2300)$ as 
the candidates of $f_4(1^3F_4)$ and $f_4(2^3F_4)$, respectively.

\begin{figure*}[htbp]
\begin{center}
\scalebox{1.82}{\includegraphics[width=\columnwidth]{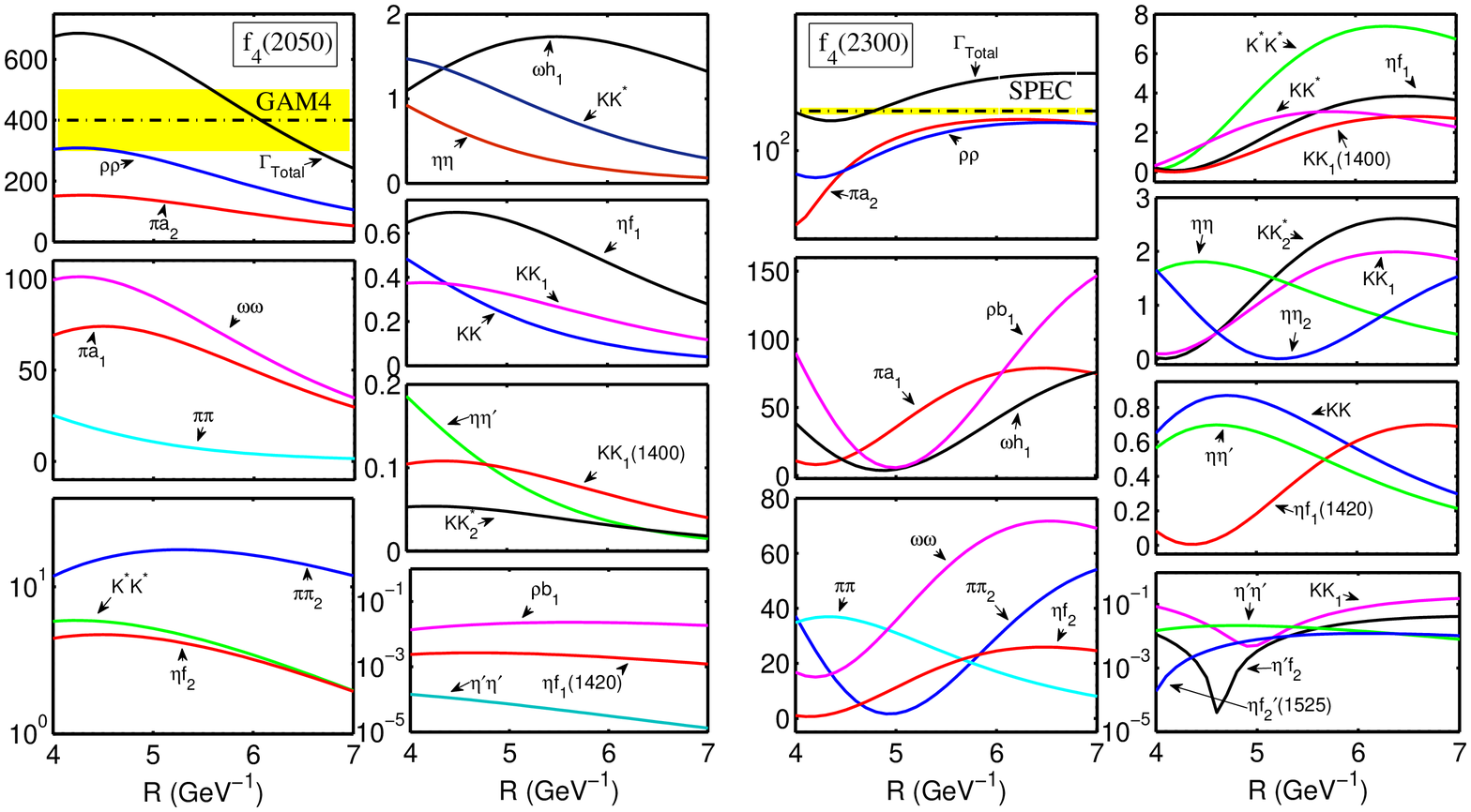}}
\caption{(color online). The $R$ dependence of the partial and total decay widths of $f_4(2050)$ and $f_4(2300)$ states.  All results are in units of MeV. The dot-dashed lines with yellow bands are experimental widths of $f_4(2050)$  \cite{Alde:1985kp} and $f_4(2300)$ \cite{Anisovich:2000ut}.} \label{f4}
\end{center}
\end{figure*}

 \begin{figure*}[htb]
\begin{center}
\scalebox{1.82}{\includegraphics[width=\columnwidth]{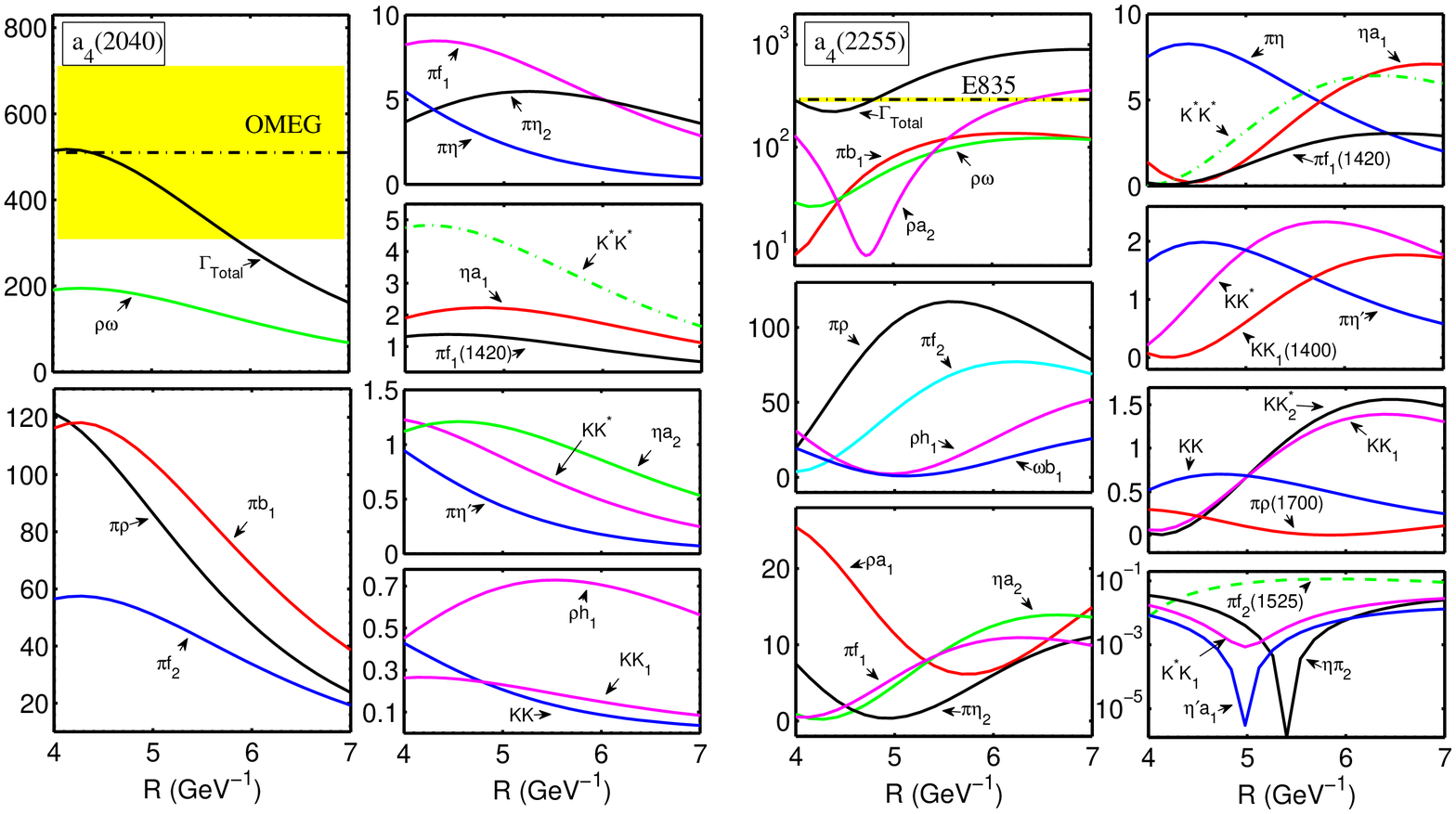}}
\caption{(color online).  The $R$ dependence of the partial and total decay widths of $a_4(2040)$ and $a_4(2255)$ states.  All results are in units of MeV. The dot-dashed lines with yellow bands are the corresponding experimental widths of  $a_4(2040)$ \cite{Corden:1977em} and $a_4(2255)$ \cite{Uman:2006xb}.} \label{a4}
\end{center}
\end{figure*}

As  isospin partners of $f_4(2050)$ and $f_4(2300)$, the decay features of $a_4(2040)$ and $a_4(2255)$ are similar to those of $f_4(2050)$ and $f_4(2300)$, respectively. Overlap exists between the experimental and calculated results of the total decay width for $a_4(2040)$  when $R= 4.0 - 5.8$ GeV$^{-1}$. The dominant  decay channels of $a_4(2040)$ are  $\rho\omega$ and $\pi b_1$ as given in Fig. \ref{a4}. The ratio ${\Gamma_{\pi\rho}}/{\Gamma_{\pi f_2}}=1.1\pm0.2\pm0.2$ was obtained in Ref. \cite{Chung:2002pu}, this can be well reproduced by our calculations, with the value $1.2 -  2.1$. In addition, we obtain the partial widths of $a_4(2040)$ decaying into $\pi\rho$ and $KK$, i.e., 
$\Gamma_{a_4(2040)\to\pi\rho
}=19 -  57$ MeV and  $\Gamma_{a_4(2040)\to KK}=0.035 - 0.43$ MeV, which deviates from the experimental data $\Gamma_{a_4(2040)\to\pi\rho
}=10\pm 3$ MeV and  $\Gamma_{a_4(2040)\to KK}=6\pm  2$ MeV, respectively in Ref. \cite{Bramon:1980xm}.  Further experimental study of $a_4(2040)$ would, thus, be useful.

In the results given in Fig. \ref{a4}, one notices the 
overlap between calculated total widths and experimental data  \cite{Uman:2006xb,Anisovich:2001pn}. Here, $\pi b_1$ and $\rho\omega$ are the main decay modes of  $a_4(2255)$, while $\pi\rho$ and $\pi f_2$ are sizable decay channels. Only $a_4(2255) \to  \pi f_2$ was reported in Ref. \cite{Bugg:2004xu}. Thus, we also suggest searching for the $\pi b_1$ and $\rho\omega$ modes for $a_4(2255)$ if  it is a $a_4(2^3F_4)$ state.

\subsubsection{$f_6(2510)$ and $a_6(2450)$}

There are two $6^{++}$ states, $f_6(2510)$ and its isospin partner $a_6(2450)$, which are treated as $1^3H_6$ states. We calculate their partial and total decay widths, presented in Fig. \ref{fig:f6a6}.

\begin{figure*}[htb]
\begin{center}
\scalebox{2}{\includegraphics[width=\columnwidth]{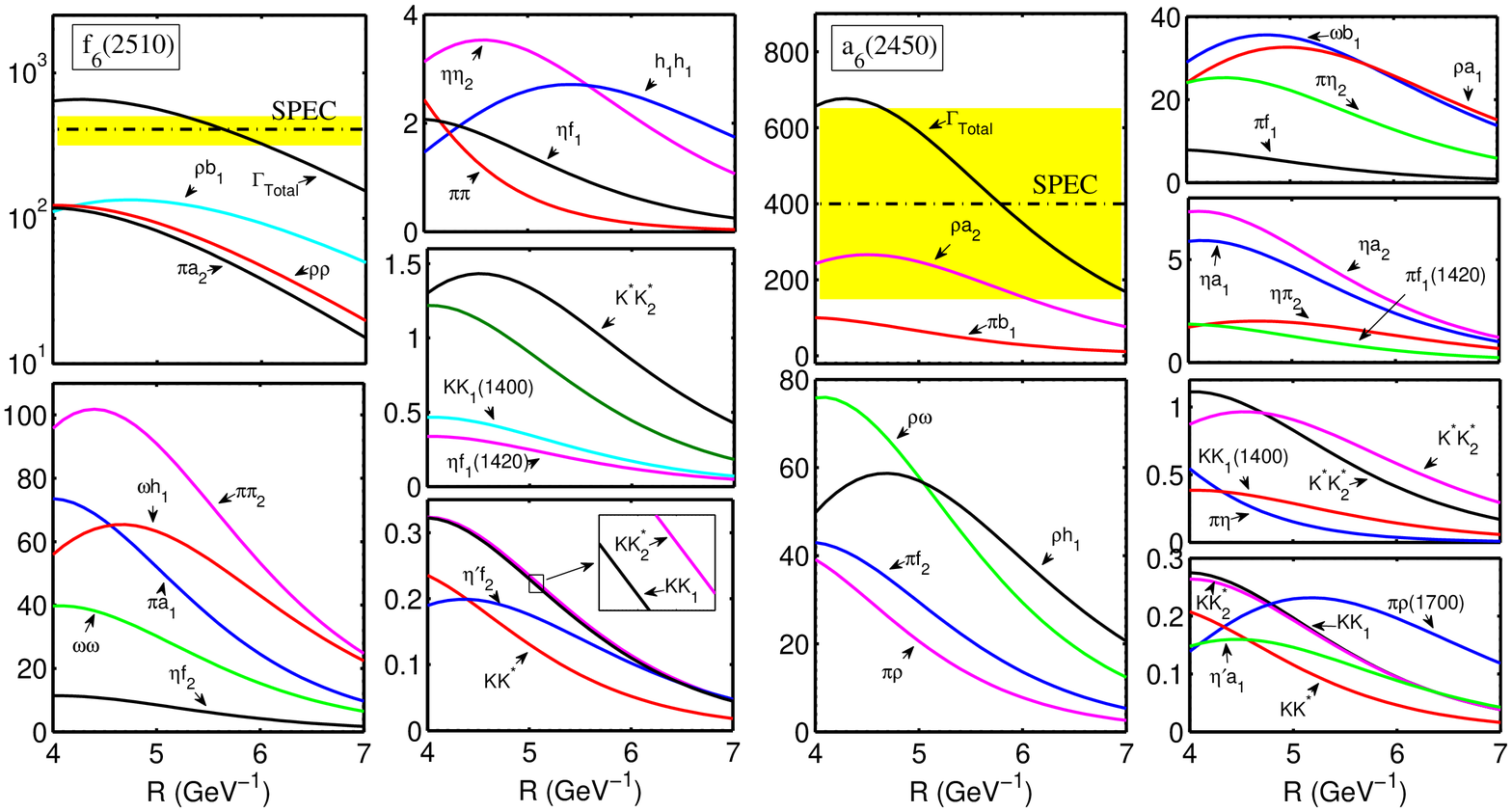}}
\caption{(color online). The $R$ dependence of the partial and total decay widths of  the $f_6(2510)$ and $a_6(2450)$ states.  All results are in units of MeV. The dot-dashed lines with yellow bands are experimental widths of $f_6(2510)$  \cite{Anisovich:2000ut} and $a_6(2450)$ \cite{Cleland:1982te}.} \label{fig:f6a6}
\end{center}
\end{figure*}

The results in Fig. \ref{fig:f6a6} show that there is an overlap between  experimental value \cite{Anisovich:2000ut} and our calculation of the total decay width of $f_6(2510)$ when $R$ in the range of $5.2 - 6.0$ GeV$^{-1}$, this gives a direct support for the $f_6(1^3H_6)$
assignment to $f_6(2510)$. Unfortunately, the  obtained  branching ratio  ${B(f_6(2510)\to \pi\pi})=3.7\times10^{- 4}  - 3.2\times10^{- 3}$ is smaller than the experimental value in Ref. \cite{Binon:1983nx} (see Table \ref{tab:1}). Since there is only one experimental measurement for this branching ratio, this ratio should be confirmed by other experiments. In addition to  the above information, we also get the dominant  decay channel of  $f_6(2510)$, i.e., $\rho b_1$ and we find that $\rho\rho$, $\pi a_2$, and $\pi\pi_2$ are its important channels.

Under the $a_6(1^3H_6)$ assignment, $a_6(2450)$ has a total decay width consistent with the  experimental  result in Ref. \cite{Cleland:1982te} if we take $R=4 - 7$ GeV$^{-1}$, where experimental data of the width has a large error bar.
The main decay channels are $\rho a_2$, $\pi b_1$, $\rho \omega$, and $\rho h_1$. The remaining OZI-allowed decay information can be found in Fig. \ref{fig:f6a6}.

\begin{figure}[htbp]
\begin{center}
\scalebox{1}{\includegraphics[width=\columnwidth]{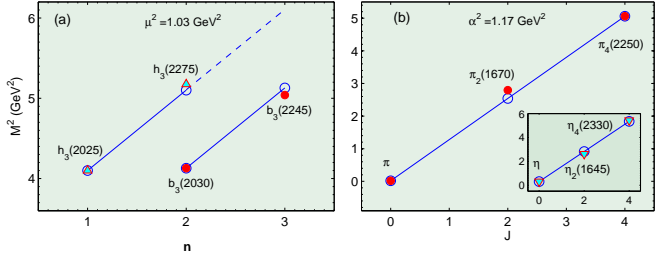}}
\caption{(color online). The analysis of Regge trajectories for the observed states with $3^{+-}$ and $J^{-+}$ quantum numbers. Diagram (a) is the $n$-$M^2$ plot for the $3^{+-}$ states, while a diagram (b) is the corresponding $J$-$M^2$ plot for the  $J^{-+}$ states. In (a), the $n$-$M^2$ trajectory with $I=1$ is
displaced one unit to the right in $n$ in order to resolve it from $I=0$.} \label{hbepnjm}
\end{center}
\end{figure}

\subsection{Four $3^{+-}$ states}

In Fig. \ref{hbepnjm} (a), we first give the $n$-$M^2$ plot analysis for four observed $3^{+-}$ states $h_3(2025)$,  $h_3(2275)$,  $b_3(2030)$, and $b_3(2245)$}. Here, $h_3(2025)$ and $h_3(2275)$
 are 
the ground state and first radial excitation in the $h_3$ meson family, while $b_3(2030)$ and $b_3(2245)$
are the isospin partners of $h_3(2025)$ and $h_3(2275)$, respectively. 

With the above assignments to the observed $3^{+-}$ states, we further discuss their strong decay behaviors.

 \begin{figure*}[htb]
\begin{center}
\scalebox{1.8}{\includegraphics[width=\columnwidth]{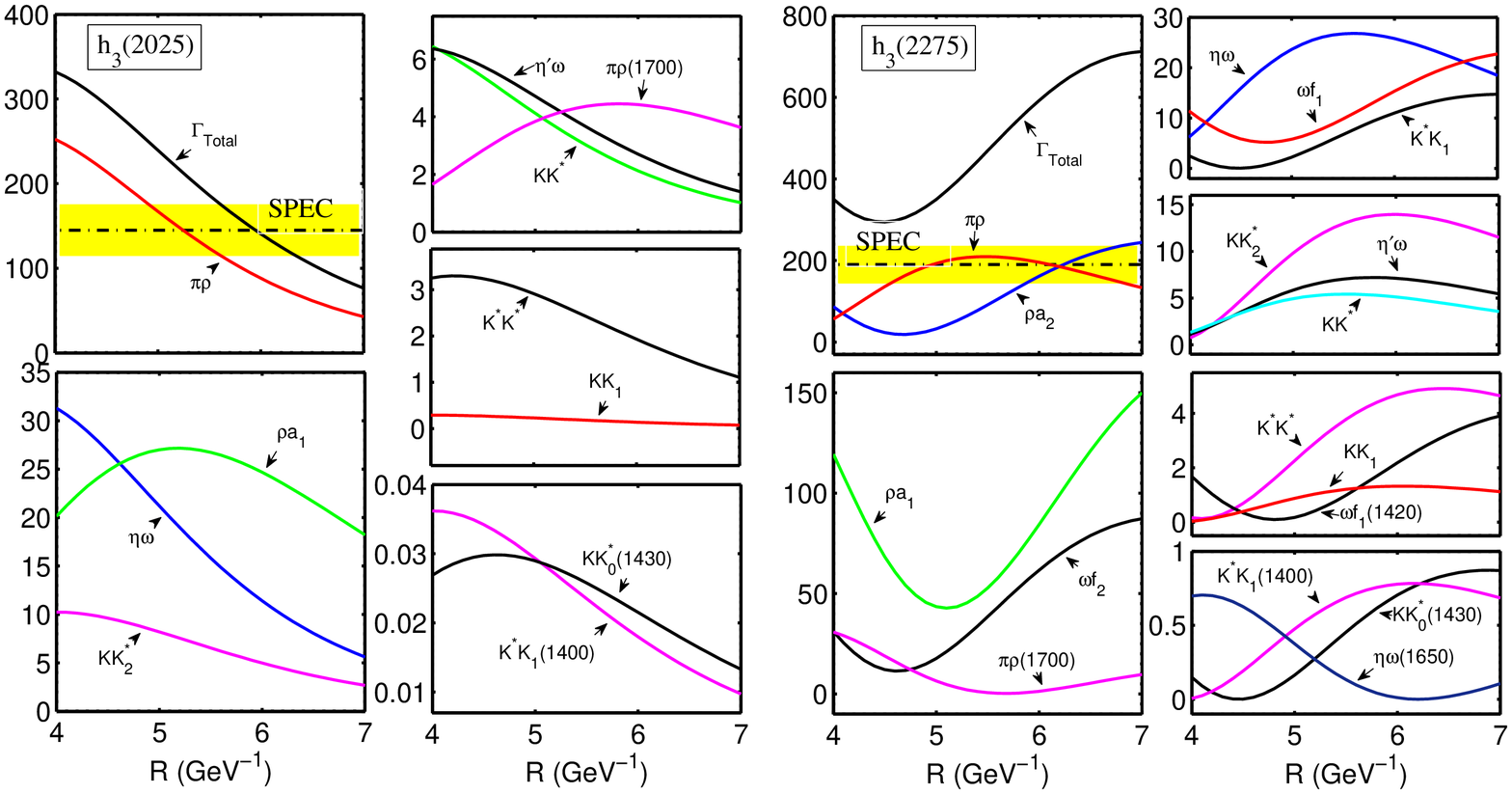}}
\caption{(color online).  The $R$ dependence of the partial and total decay widths of $h_3(2025)$ and $h_3(2275)$ states.  All results are in units of MeV. The dot-dashed lines with yellow bands are the corresponding experimental widths of  $h_3(2025)$ and $h_3(2275)$ \cite{Anisovich:2011sva}.} \label{h3}
\end{center}
\end{figure*}

 \begin{figure*}[htb]
\begin{center}
\scalebox{1.8}{\includegraphics[width=\columnwidth]{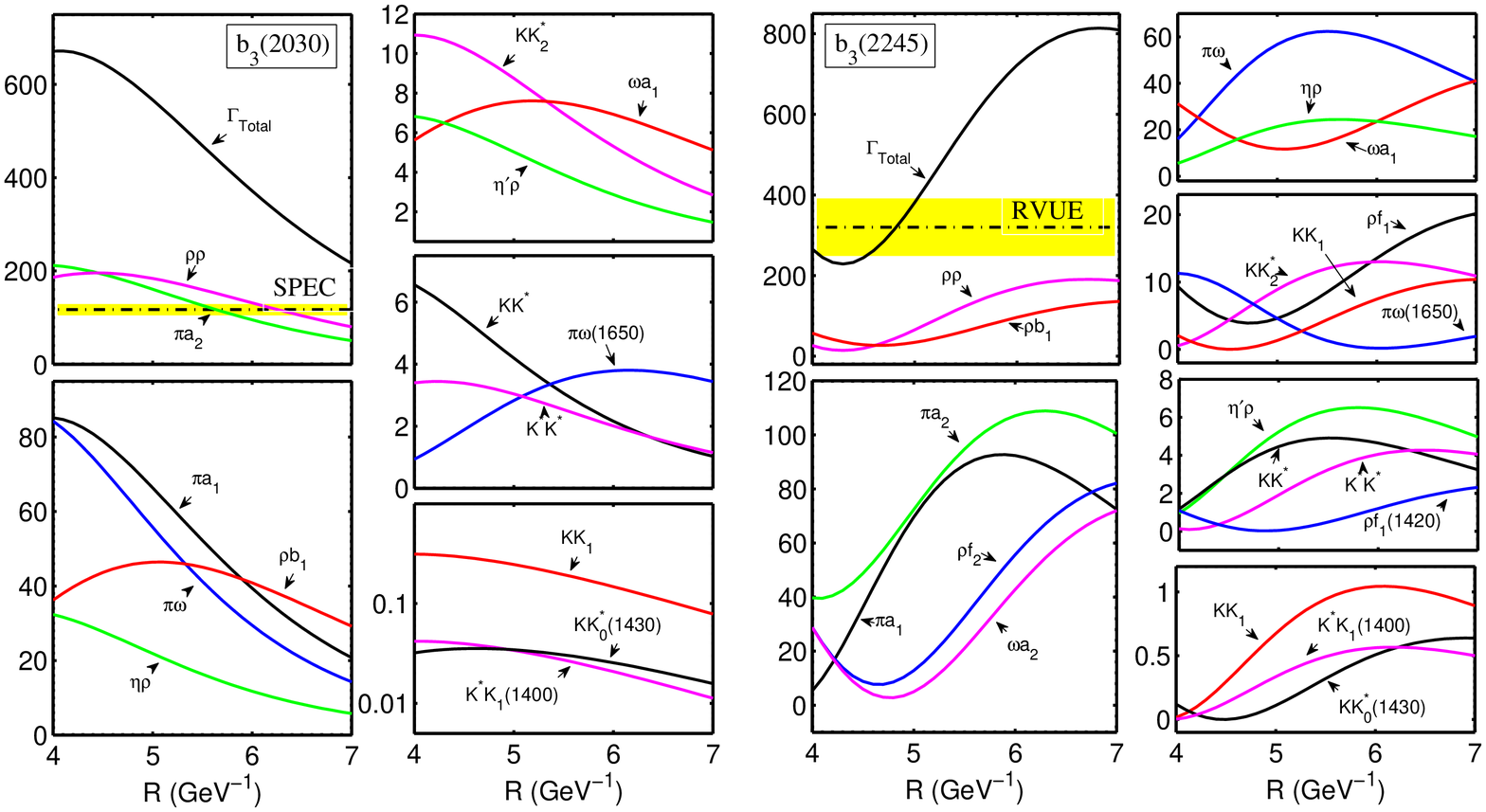}}
\caption{(color online).  The $R$ dependence of the partial and total decay widths of $b_3(2030)$ and $b_3(2245)$ states.  All results are in units of MeV. The dot-dashed lines with yellow bands are the corresponding experimental widths of  $b_3(2030)$ \cite{Anisovich:2002su}, and $b_3(2245)$ \cite{Bugg:2004xu}.} \label{b3}
\end{center}
\end{figure*}

As illustrated in Fig. \ref{h3}, $\pi \rho $, $\eta\omega$, and $\rho a_1$ are the main decay channels of  $h_3(2025)$, which can explain why this  was observed in the $\eta\omega$ channel. In addition, we find that a theoretical result overlaps with the experimental width \cite{Anisovich:2011sva}. Thus, we suggest future experimental study of $h_3(2025)$ by its other dominant decay modes $\pi\rho$ and $\rho a_1$, which are still missing in experiment. 

$h_3(2275)$ mainly decays into $\rho a_2$,  $\pi\rho$, $\rho a_1$, $\omega f_2$, and $\eta\omega$. This state was observed in the processes $p\bar{p}\to\eta\omega, \omega\pi^0\pi^0$ \cite{Anisovich:2011sva}. The detailed decay information of $h_3(2275)$  can be found in Fig. \ref{h3}. We notice that  inconsistency  exists inconsistency between the experimental width and the obtained total decay width, since the calculated total decay width under $R=4 - 7$ GeV$^{-1}$ is larger than the experimental data \cite{Anisovich:2011sva}.

From the results in Fig. \ref{b3}, we conclude that the main decay modes of $b_3(2030)$ are $\pi a_2$ and   $\rho\rho$, and  $\pi a_1$, $\pi\omega$,  $\rho b_1$, and $\eta\rho$ have sizable contributions to the total decay width. Here, 
$\pi\omega$ and $\pi^+\pi^-$ decay channels of $b_3(2030)$ were observed in experiment \cite{Anisovich:2002su}, where $\pi^+\pi^-$ can be from $\rho$. This experimental phenomenon does not contradict  our theoretical result. However, the obtained total decay width of $b_3(2030)$ cannot fall into the range of experimental width when taking $R=4 - 7$ GeV$^{-1}$, which is a situation similar to 
$h_3(2275)$. In future,  we need more experimental measurements of the resonance parameters of  $h_3(2275)$ and $b_3(2030)$. 

Another observed $b_3$ state is  $b_3(2245)$. As displayed in Fig. \ref{b3}, there is an overlap between calculated total width and experimental data \cite{Bugg:2004xu}. Its main decay channels are $\rho\rho$, $\rho b_1$, and $\pi a_2$; the  
$\pi a_1$, $\rho f_2$, $\omega a_2$ and $\pi\omega$ channels have important contributions to the total decay width, where $\omega a_2$ and $\pi\omega$ are the observed channels \cite{Bugg:2004xu}.

\subsection{Two $4^{-+}$ states}

In this subsection, we discuss the last two observed high-spin states, $\eta_4(2330)$ and $\pi_4(2250)$, which have $4^{-+}$  quantum numbers (see Table \ref{tab:1}).
The corresponding analysis of Regge trajectories with the $J$-$M^2$ plot is shown in Fig. \ref{hbepnjm} (b) ; this was used to study  $2^{-+}$ states in our previous work \cite{Wang:2014sea}. 
 $\pi$ and $\pi_2$ \cite{Wang:2014sea} are the ground states of their own families. Thus, Fig. \ref{hbepnjm} (b) indicates that $\eta_4(2330)$ and $\pi_4(2250)$ are the ground states of the $\eta_4$ and $\pi_4$ meson families. In fact, Refs. \cite{Anisovich:2000kxa,Klempt:2007cp,Afonin:2007aa,Anisovich:2002us} gave the same suggestion. In the following, we calculate their two-body strong decays with the assignments $\eta_4(1^1G_4)$ and $\pi_4(1^1G_4)$ to $\eta_4(2330)$ and $\pi_4(2250)$, respectively.

Our calculated theoretical total width of $\eta_4(2330)$ is larger than  experimental data \cite{Anisovich:2000ut}, where the main decay channels are $\rho b_1$, $\rho\rho$ and $\pi a_2$, while $\pi a_1$, $\omega h_1$,  $\omega\omega$, and $\eta f_2$ are its important decay channels. $\eta_4(2330)$ 
was  first reported in the final states $(\pi a_2)_{L=4}$ and $(a_0\pi)_{L=4}$, and was also observed in the $\eta f_2$ channel \cite{Bugg:2004xu}. The information from its partial decay width shows that $\eta_4(2330)$  as $1^1G_4$ is reasonable. At present,  a crucial task is to further check 
the resonance parameters of $\eta_4(2330)$. 

\begin{figure*}[htb]
\begin{center}
\scalebox{2}{\includegraphics[width=\columnwidth]{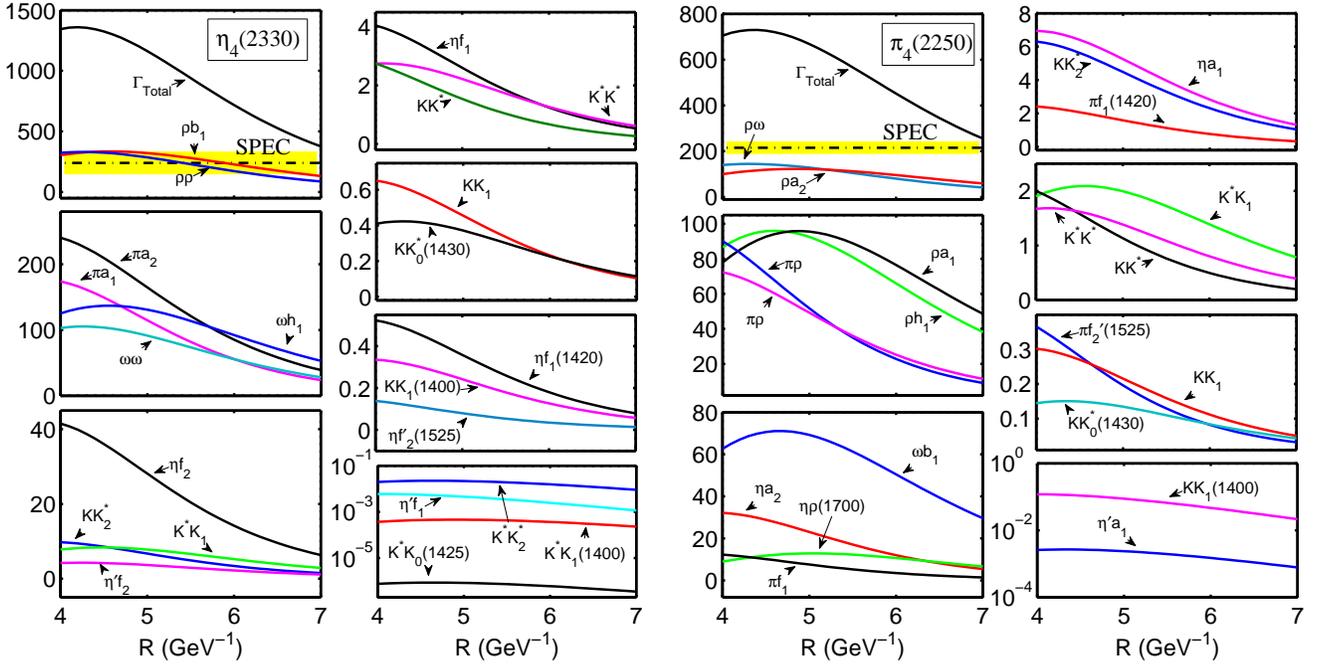}}
\caption{(color online).  The $R$  dependence of the and total decay widths of $\pi_4(2250)$ and $\eta_4(2330)$ states.  All results are in units of MeV. The dot-dashed lines with yellow bands are the corresponding experimental widths of  $\pi_4(2250)$ \cite{Anisovich:2011sva} and $\eta_4(2330)$ \cite{Anisovich:2000ut}.} \label{eta4pi4}
\end{center}
\end{figure*}

Figure. \ref{eta4pi4}  present the decays of $\pi_4(2250)$. 
We find the theoretical total width  is larger than the SPEC data \cite{Anisovich:2001pn} if we take the $R=4 - 7$ GeV$^{-1}$ range.  
$\pi_4(2250)$ mainly decays into $\rho \omega$, $\rho a_2$, $\rho h_1$ and $\rho a_1$.

{Before closing this section, we list some additional important ratios in Table \ref{tab:sumratio}, where we collect the corresponding $R$ values that  can be adopted to reproduce the experimental data.} 
  \renewcommand{\arraystretch}{1.5}
\begin{table*}[htbp]
\caption{The typical branching ratios of  these discussed high-spin mesons corresponding to the successful $R$ values.  \label{tab:sumratio}}
\begin{center}
\begin{tabular}{ccc}
\toprule[1pt]
\midrule[1pt]
States&                             $R$ (GeV$^{-1}$)&                        Ratios  \\
\midrule[1pt]

 $a_3(1875)$  & 4.0 $ - $ 4.6
 &  $\left.{\Gamma _{\pi f_2}}/{\Gamma _{\text{Total}}} \right.= 0.33 -  0.34,
~\left.{\Gamma _{\pi \rho }}/{\Gamma _{\text{Total}}} \right. = 0.3  -  0.33,
~\left.{\Gamma _{\pi \rho }}/{\Gamma _{\pi f_2}} \right. = 0.91 -  0.98,
~{\Gamma _{\rho \omega }}/{\Gamma _{\text{Total}}} = 0.22  -  0.27$ \\

 $a_3(2030)$  &4.0 $ - $ 7.0& $
 \left.{\Gamma _{\rho \omega }}/{\Gamma _{\text{Total}}} \right. = 0.26  - 0.32,
~\left.{\Gamma _{\rho h_1}}/{\Gamma _{\rho \omega }} \right. = 0.53  -  0.75,
~ \left.{\Gamma _{\pi f_2}}{\Gamma _{\pi \rho }} \right. = 0.41  - 0.51 ,
~\left.{\Gamma _{\eta a_2}}/{\Gamma _{\text{Total}}} \right. = 0.065  -  0.087$\\
$a_3(2275)$   &   4.6 $-$ 5 &
$
\left.{\Gamma _{\rho a_1}}/{\Gamma _{\text{Total}}} \right. = 0.066 - 0.077,
~ \left.{\Gamma _{\pi \rho }}/{\Gamma _{\text{Total}}} \right. = 0.23  -  0.27,
~ \left.{\Gamma _{\pi \rho }}/{\Gamma _{\rho a_1}} \right. = 3.3 - 3.8,
~ \left.{\Gamma _{\rho h_1}}/{\Gamma _{\text{Total}}}  \right.= 0.075 -  0.099$ \\

 $\omega_3(1670)$ &4.0 $ -$  5.4 &$
\left.{\Gamma _{\pi \rho }}/{\Gamma _{\text{Total}}} \right.  = 0.69  -  0.75,
~ \left.{\Gamma _{\pi b_1}}/{\Gamma _{\text{Total}}} \right.  = 0.2  -  0.26,
~\left.{\Gamma _{\eta \omega }}/{\Gamma _{\text{Total}}}  \right. = 0.035  -  0.036,$\\
&&$
\left.{\Gamma _{\eta \omega }}/{\Gamma _{\pi \rho }} \right.  =0.046  -  0.051,
~\left.{\Gamma _{\eta \omega }}/{\Gamma _{\pi b_1}} \right.  = 0.13  -  0.17,
~ \left.{\Gamma _{KK^*}}/{\Gamma _{\text{Total}}} \right.  = 0.0028  -  0.0032$ \\

$\omega_3(1945)$  &5.3  $-$  7.0
 &$\left.{\Gamma _{\pi \rho }}/{\Gamma _{\text{Total}}} \right.= 0.81  - 0.84,
~\left.{\Gamma _{\pi b_1}}/{\Gamma _{\text{Total}}} \right.= 0.073  -  0.092,
~ \left.{\Gamma _{\pi b_1}}{\Gamma _{\pi \rho }} \right.= 0.087  - 0.11,$ \\
&&$
\left.{\Gamma _{\eta \omega }}/{\Gamma _{\text{Total}}}  \right.= 0.05 - 0.056,
~ \left.{\Gamma _{\eta \omega }}/{\Gamma _{\pi \rho }} \right.= 0.059  - 0.068,
~\left.{\Gamma _{\eta \omega }}/{\Gamma _{\pi b_1}} \right.= 0.6  -  0.69$\\

 $\phi_3(1850)$ &5.3 $ - $ 7.0 &$
 \left.{\Gamma _{K^*K^*}}/{\Gamma _{\text{Total}}} \right. = 0.59 -  0.65,
~ \left.{\Gamma _{KK^*}}/{\Gamma _{\text{Total}}}  \right.= 0.24 -  0.27,
~\left.{\Gamma _{KK^*}}/{\Gamma _{K^*K^*}}  \right.= 0.36 -  0.46$ \\
$\omega_3(2285)$ &4.7  $-$  5.0&$
\left.{\Gamma _{\pi \rho }}/{\Gamma _{\text{Total}}}  \right.= 0.7  \text{--} 0.71, 
~\left.{\Gamma _{\pi b_1}}{\Gamma _{\text{Total}}}  \right.= 0.13  \text{--}0.13,
~\left.{\Gamma _{\pi b_1}}/{\Gamma _{\pi \rho }} \right. = 0.18  \text{--}0.19,
~\left.{\Gamma _{\eta \omega }}/{\Gamma _{\text{Total}}} \right. = 0.059 \text{--} 0.065$, \\
&&
$\left.{\Gamma _{\eta \omega }}/{\Gamma _{\pi \rho }} \right. = 0.083  \text{--}  0.092, 
~\left.{\Gamma _{\eta \omega }}/{\Gamma _{\pi b_1}} \right. = 0.47  \text{--}0.49, 
~\left.{\Gamma _{\rho a_1}}/{\Gamma _{\text{Total}}} \right. = 0.019  \text{--}  0.031, 
~\left.{\Gamma _{\rho a_1}}/{\Gamma _{\pi \rho }} \right.= 0.027\text{--} 0.044$ \\

$\omega_3(2255)$  &6.2  $-$  7.0&
$\left.{\Gamma _{\rho a_1}}/{\Gamma _{\text{Total}}} \right.= 0.27  \text{--} 0.36,
~\left.{\Gamma _{\rho a_2}}/{\Gamma _{\rho a_1}} \right.=0.68  \text{--}0.84, 
~\left.{\Gamma _{\omega f_2}}/{\Gamma _{\text{Total}}} \right.=0.088 \text{--}0.14, 
~\left.{\Gamma _{\omega f_2}}/{\Gamma _{\rho a_1}} \right.= 0.32\text{--}  0.38$, \\
&&
$\left.{\Gamma _{\omega f_2}}/{\Gamma _{\rho a_2}} \right. =0.45 \text{--} 0.48,
  ~\left.{\Gamma _{\pi b_1}}/{\Gamma _{\pi \rho }} \right.  = 0.33  \text{--}  0.47,
~\left.{\Gamma _{\pi \rho (1700)}}/{\Gamma _{Total}} \right. = 0.043  \text{--}  0.052, 
 ~\left.{\Gamma _{\pi \rho (1700)}}/{\Gamma_{\rho a_1}} \right. = 0.14 \text{--}  0.16$ \\

$f_4(2050)$ &4.0 $-$7.0&$
 \left.{\Gamma _{\rho \rho }}/{\Gamma _{\text{Total}}}\right.  = 0.43 - 0.45 ,
~ \left.{\Gamma _{\pi a_2}}/{\Gamma _{\text{Total}}}\right.  = 0.22,
~ \left.{\Gamma _{\pi a_2}}/{\Gamma _{\rho \rho }} \right. = 0.49 - 0.5$\\ 
&&
$ \left.{\Gamma _{\omega \omega }}/{\Gamma _{\text{Total}}}\right.  = 0.14- 0.15,
~ \left.{\Gamma _{\omega \omega }}/{\Gamma _{\rho \rho }}\right.  = 0.33,
~ \left.{\Gamma _{\omega \omega }}/{\Gamma _{\pi a_2}} \right. =  0.66 $\\

 $f_4(2300)$  &  4.0 $ -$  7.0
 &$\left.{\Gamma _{\pi a_2}}/{\Gamma _{\text{Total}}} \right.= 0.020  -  0.25,
~ \left.{\Gamma _{\rho \rho }}/{\Gamma _{\text{Total}}} \right.= 0.092  -  0.21,
~ \left.{\Gamma _{\rho \rho }}/{\Gamma _{\pi a_2}} \right.= 0.77  -  4.3,
~ \left.{\Gamma _{\pi a_1}}/{\Gamma _{\text{Total}}}\right. = 0.020  -  0.075$\\

 $a_4(2040)$ &4.0 $-$ 5.8 &$
\left.{\Gamma _{\rho \omega }}/{\Gamma _{\text{Total}}} \right.= 0.36 - 0.38,
~ \left.{\Gamma _{\pi \rho }}/{\Gamma _{\text{Total}}}\right. = 0.16 - 0.23,
~ \left.{\Gamma _{\pi \rho }}/{\Gamma _{\rho \omega }} \right.= 0.42  -  0.64,
~ \left.{\Gamma _{\pi b_1}}/{\Gamma _{\text{Total}}} \right.= 0.22$,\\
&&
$\left.{\Gamma _{\pi b_1}}/{\Gamma _{\rho \omega }} \right.= 0.59  -  0.61,
~ \left.{\Gamma _{\pi b_1}}/{\Gamma _{\pi \rho }} \right.= 0.96 - 1.4,
~ \left.{\Gamma _{\pi f_2}}/{\Gamma _{\text{Total}}}\right. = 0.11,
~ \left.{\Gamma _{\pi f_2}}/{\Gamma _{\rho \omega }}\right. = 0.29 - 0.3$\\
$a_4(2255)$   &  4.7  $-$  4.8 &  
$ \left.{\Gamma _{\pi f_2}}/{\Gamma _{\pi b_1}} \right.= 0.41  -  0.57 ,
~ \left.{\Gamma _{\pi f_1}}/{\Gamma _{\pi f_2}} \right.= 0.092  -  0.14 ,
~\left.{\Gamma _{\text{$\pi \eta $'}}}/{\Gamma _{\pi \eta }} \right.= 0.22  -  0.29$\\
 $\pi_4(2250)$ & 4.0 $ -$  7.0&
$ \left.{\Gamma _{\rho \omega }}/{\Gamma _{\text{Total}}} \right.= 0.15  -  0.19 .
~\left.{\Gamma _{\rho a_2}}/{\Gamma _{\text{Total}}}\right.= 0.14  -  0.23 ,
~ \left.{\Gamma _{\rho h_1}}/{\Gamma _{\text{Total}}}\right.=0.12  -  0.15 ,
~ \left.{\Gamma _{\rho h_1}}/{\Gamma _{\rho \omega }}\right.= 0.62  - 0.92 $,\\
&&$
 \left.{\Gamma _{\rho h_1}}/{\Gamma _{\rho a_2}} \right.= 0.64  -  0.87,
~ \left.{\Gamma _{\rho a_1}}/{\Gamma _{\rho a_2}}\right.= 0.78  -  0.82,
~\left.{\Gamma _{\rho a_1}}/{\Gamma _{\rho h_1}} \right.= 0.9  -  1.3 ,
~ \left.{\Gamma _{\pi f_2}}/{\Gamma _{\pi \rho }}\right.= 0.8  -  1.3$\\

$\eta_4(2330)$&  4.0  $-$  7.0
&$ \left.{\Gamma _{\rho b_1}}/{\Gamma _{\text{Total}}}\right. = 0.23  -  0.35,
~ \left.{\Gamma _{\rho \rho }}/{\Gamma _{\text{Total}}}\right. = 0.23  -  0.25,
~ \left.{\Gamma _{\pi a_2}}/{\Gamma _{\rho \rho }}\right. = 0.45  -  0.74,
~ \left.{\Gamma _{\pi a_1}}/{\Gamma _{\pi a_2}} \right.= 0.61  -  0.72,$\\
&&$
 \left.{\Gamma _{\omega h_1}}/{\Gamma _{\text{Total}}}\right. = 0.093  -  0.14,
~ \left.{\Gamma _{\omega h_1}}/{\Gamma _{\rho b_1}} \right.= 0.4  - 0.41,
~ \left.{\Gamma _{\omega h_1}}/{\Gamma _{\rho \rho }}\right. = 0.39  -  0.61,
~ \left.{\Gamma _{\omega \omega }}/{\Gamma _{\text{Total}}}\right. =0.075  -  0.079 $\\

 $\rho_4(2230)$&6.8$-$7.0  & $
 \left.{\Gamma _{\rho \rho }}/{\Gamma _{\text{Total}}} \right.= 0.2  -  0.21,
~ \left.{\Gamma _{\rho b_1}}/{\Gamma _{\text{Total}}} \right.= 0.19  -  0.27,
~ \left.{\Gamma _{\rho b_1}}/{\Gamma _{\rho \rho }}\right. = 0.92  -  1.3 ,
~ \left.{\Gamma _{\pi a_1}}/{\Gamma _{\text{Total}}}\right. = 0.094  -  0.12,$\\
&&$
 \left.{\Gamma _{\pi a_1}}/{\Gamma _{\rho \rho }}\right. = 0.46  -  0.60,
~ \left.{\Gamma _{\rho f_2}}/{\Gamma _{\rho b_1}} \right.= 0.3  -  0.39,
~ \left.{\Gamma _{\omega a_1}}/{\Gamma _{\rho b_1}} \right.= 0.29  -  0.33$ \\
$\omega_4(2250)$  &4.0$ -$ 7.0&$
 \left.{\Gamma _{\rho a_1}}/{\Gamma _{\text{Total}}}\right. = 0.27  -  0.36,
~  \left.{\Gamma _{\rho a_2}}/{\Gamma _{\rho a_1}} \right.= 0.68 - 0.84,
~ \left.{\Gamma _{\omega f_2}}/{\Gamma _{Total}} \right.= 0.088 - 0.14,
~ \left.{\Gamma _{\omega f_2}}/{\Gamma _{\rho a_1}} \right.= 0.32 - 0.38, 
 $\\
&&$
 \left.{\Gamma _{\omega f_2}}/{\Gamma _{\rho a_2}}\right.= 0.45- 0.48,
~ \left.{\Gamma _{\pi b_1}}/{\Gamma _{\pi \rho }} \right.= 0.33- 0.47,
~\left.{\Gamma _{\pi \rho (1700)}}/{\Gamma _{Total}}\right. = 0.043 -0.052 ,
~ \left.{\Gamma _{\pi \rho (1700)}}/{\Gamma _{\rho a_1}}\right.= 0.14 - 0.16$\\
\midrule[1pt]

$\omega_5(22250)$& 4.0$-$7.0  &$
  \left.{\Gamma _{\rho a_2}}/{\Gamma _{\text{Total}}}\right. = 0.42 - 0.58,
~ \left.{\Gamma _{\omega f_2}}/{\Gamma _{\text{Total}}} \right.= 0.19 - 0.23,
~ \left.{\Gamma _{\omega f_2}}/{\Gamma _{\rho a_2}} \right.= 0.4 - 0.45,
~ \left.{\Gamma _{\pi \rho }}/{\Gamma _{\pi b_1}}\right. = 0.35 - 0.57,$
\\&&
$ \left.{\Gamma _{\eta h_1}}/{\Gamma _{\pi b_1}} \right.= 0.1 - 0.12,
~ \left.{\Gamma _{\eta \omega }}/{\Gamma _{\pi b_1}} \right.= 0.042 - 0.058,
~ \left.{\Gamma _{\eta \omega }}/{\Gamma _{\pi \rho }}\right. = 0.1 - 0.12,
~ \left.{\Gamma _{\eta \omega }}/{\Gamma _{\eta h_1}} \right.= 0.35 - 0.58$\\

 $\rho_5(2350)$ &4.0  $-$  7.0&$
 \left.{\Gamma _{\rho f_2}}/{\Gamma _{\text{Total}}}\right. = 0.20 -  0.21,
~ \left.{\Gamma _{\omega a_2}}/{\Gamma _{\text{Total}}}\right. = 0.17 -  0.24,
~ \left.{\Gamma _{\omega a_2}}/{\Gamma _{\rho f_2}} \right.= 0.86 -  1,
~ \left.{\Gamma _{\rho \rho }}/{\Gamma _{\text{Total}}}\right. = 0.087 -  0.12,$\\
&&$
 \left.{\Gamma _{\pi a_2}}/{\Gamma _{\rho \rho }}\right. = 0.82 -  0.99,
~ \left.{\Gamma _{\pi h_1}}/{\Gamma _{\rho \rho }}\right.= 0.53 -  0.78,
~ \left.{\Gamma _{\pi h_1}}/{\Gamma _{\pi a_2}}\right. = 0.64 -  0.78,
~ \left.{\Gamma _{\rho b_1}}/{\Gamma _{\rho f_2}} \right.= 0.33 -  0.54$ \\

\midrule[1pt]
 $f_6(2510)$ & 6.0 $-$ 6.4 & $
 \left.{\Gamma _{\rho b_1}}/{\Gamma _{\text{Total}}}\right.= 0.28  -  0.3,
~ \left.{\Gamma _{\rho \rho }}/{\Gamma _{\text{Total}}} \right.= 0.14  -  0.14,
~ \left.{\Gamma _{\rho \rho }}/{\Gamma _{\rho b_1}}\right. = 0.46  -  0.51,
~ \left.{\Gamma _{\pi a_2}}/{\Gamma _{\text{Total}}}\right. = 0.11  -  0.12,$\\
&&$\
 \left.{\Gamma _{\pi a_2}}/{\Gamma _{\rho b_1}}\right. = 0.36  -  0.41,
~ \left.{\Gamma _{\pi a_2}}/{\Gamma _{\rho \rho }}\right. = 0.79  -  0.81,
~ \left.{\Gamma _{\pi \pi _2}}/{\Gamma _{\text{Total}}}\right. = 0.16,
~ \left.{\Gamma _{\pi \pi _2}}/{\Gamma _{\rho b_1}} \right.= 0.54  -  0.58$ \\

 $a_6(2450)$  &4.0 $-$ 7.0&$
 \left.{\Gamma _{\rho a_2}}/{\Gamma _{\text{Total}}} \right.=0.37 -  0.46,
~ \left.{\Gamma _{\rho \omega }}/{\Gamma _{\text{Total}}} \right.= 0.074  -  0.12,
~ \left.{\Gamma _{\rho \omega }}/{\Gamma _{\pi b_1}} \right.= 0.75 -  1.1,
~ \left.{\Gamma _{\rho h_1}}/{\Gamma _{\text{Total}}}\right. = 0.076  - 0.12,$ \\
&&$
 \left.{\Gamma _{\rho h_1}}/{\Gamma _{\rho a_2}} \right.= 0.21 -  0.27,
~ \left.{\Gamma _{\pi f_2}}/{\Gamma _{\pi b_1}} \right.= 0.43 -  0.48,
~ \left.{\Gamma _{\pi f_2}}/{\Gamma _{\rho \omega }} \right.= 0.43 -  0.57,
~ \left.{\Gamma _{\pi \rho }}/{\Gamma _{\pi b_1}}\right.= 0.24 - 0.39$ \\
\bottomrule[1pt]
\bottomrule[1pt]
\end{tabular}
\end{center}
\end{table*}

\section{Conclusions and discussion}\label{sec4}

In this work, we have mainly focused on the study of 26 high-spin states reported in experiments; we have performed the mass spectrum analysis and have carried out the calculation of their two-body OZI-allowed strong decays, which is helpful in revealing their underlying features. The first task is to explore whether the observed high-spin states can be categorized into conventional meson families.

The analysis of Regge trajectories with the $n$-$M^2$ and $J$-$M^2$ plots has provided an effective approach to  study the meson categorization phenomenologically. We have discussed the possible meson assignments to the observed high-spin states listed in PDG \cite{Agashe:2014kda}. The main task of the present work has been the calculation of the two-body OZI-allowed strong decays of the high-spin states, which can be applied to test the possible meson assignments. In Sec. \ref{sec3}, we have discussed this point in detail. The predicted decay behaviors of the  discussed high-spin states  can provide valuable information for  further experimental study in the future. 

At present, most of the high-spin states reported in experiments are collected into the further states in the PDG \cite{Agashe:2014kda}, because the experimental information of these high-spin unflavored states is not abundant. Thus, we suggest that more experimental measurements of the resonance parameters should be obtained , and the missing main decay channels searched for. Such efforts will be helpful in establishing these high-spin states in experiments. 

With the experimental progress, the exploration of high-spin mesons is becoming an important issue in hadron physics, with good  platforms in  BESIII, BelleII, and COMPASS experiments. We hope that inspired by this work,  more experimental and theoretical studies of high-spin states are conducted in the future.

\section*{Acknowledgments}

This project is supported by the National Science Foundation for Fostering Talents in Basic Research of the National Natural Science Foundation of China and the National Natural Science Foundation of China under Grants No. 11222547 and No. 11175073, the Ministry of Education of China (SRFDP under Grant No. 2012021111000), and the Fok Ying Tung Education Foundation (Grant No. 131006).


\end{document}